\documentclass[11pt,noindent]{article}
\usepackage{graphics,cite,amssymb,epsfig,float}
\usepackage{amsmath, mathbbol}
\usepackage{amsfonts,xfrac,slashed}
\usepackage{a4,graphicx}
\usepackage{fancybox}

\usepackage{amsfonts}
\usepackage{ulem}
\usepackage{soul}
\usepackage{bm}

\usepackage{xcolor,color}
\definecolor{cblue}{RGB}{100,5,255}
\definecolor{cred}{RGB}{180,50,40} 
\definecolor{cgreen}{RGB}{40,255,40} 
\definecolor{corange}{RGB}{180,140,20} 
\definecolor{cmagenta}{RGB}{120,20,50} 
\definecolor{mathblue}{RGB}{100,100,250}

\usepackage{array}
\usepackage{multirow}
\def\lsim{\;\raise0.3ex\hbox{$<$\kern-0.75em\raise-1.1ex\hbox{$\sim$}}\;}
\def\gsim{\;\raise0.3ex\hbox{$>$\kern-0.75em\raise-1.1ex\hbox{$\sim$}}\;}

\def\ben{\begin{enumerate}}  \def\een{\end{enumerate}}
\def\bit{\begin{itemize}}    \def\eit{\end{itemize}}
\def\beq{\begin{equation}}   \def\eeq{\end{equation}}
\def\ba{\begin{array}}       \def\ea{\end{array}}
\def\bea{\begin{eqnarray}}   \def\eea{\end{eqnarray}}

\newcolumntype{C}{ >{\centering\arraybackslash} m{4cm} }
\usepackage{hhline}

\begin{document}

\setcounter{footnote}{0}
\vspace*{-1.5cm}
\begin{flushright}
{LPT-Orsay-18-86}
\end{flushright}
\begin{center}
\vspace*{1mm}

\vspace{1cm}
{\Large\bf 
Heavy neutral leptons and high-intensity observables
}

\vspace*{0.8cm}

{\bf Asmaa Abada$^{a}$ and Ana M. Teixeira$^{b}$}  
   
\vspace*{.5cm} 
$^{a}$ Laboratoire de Physique Th\'eorique (UMR 8627), CNRS, \\
Univ. Paris-Sud, Universit\'e Paris-Saclay, 91405 Orsay, France
\vspace*{.2cm}

$^{b}$ Laboratoire de Physique de Clermont (UMR 6533), CNRS/IN2P3,\\ 
Univ. Clermont Auvergne, 4 Av. Blaise Pascal, F-63178 Aubi\`ere Cedex, France
\end{center}

\vspace*{6mm}
\begin{abstract}
New Physics models in which the Standard Model particle content is
enlarged via the addition of sterile fermions remain among the  
most minimal and yet most appealing constructions, particularly since
these states are present as building blocks of numerous
mechanisms of neutrino mass generation. 
Should the new sterile states have non-negligible
mixings to the active (light) neutrinos, and if they are not 
excessively heavy,
one expects important  
contributions to numerous high-intensity observables, among them
charged lepton flavour violating 
muon decays and transitions, and lepton electric dipole moments.  
We briefly review the prospects of these minimal SM extensions to
several of the latter observables, considering both simple extensions  
and complete models of neutrino mass generation. We emphasise the
existing synergy between different observables at the Intensity   
Frontier, which will be crucial in unveiling the new model at work.
\end{abstract}
\vspace*{20mm}

\section{Introduction}
Several observational problems fuel the need to
extend the Standard Model (SM): among them, the baryon asymmetry of
the Universe (BAU), the absence of a dark matter candidate, and
neutrino oscillation phenomena (i.e. neutrino masses and mixings).
Many well-motivated New Physics (NP) scenarios have been proposed to
overcome the observational (and theoretical) caveats of the SM: the 
beyond the Standard Model (BSM) constructions either rely on extending
the particle content, enlarging the symmetry group, or then embedding
the SM into larger frameworks. Interestingly, a common ingredient of
many of the previously mentioned possibilities is the presence of
additional neutral leptons, sterile states (singlets under the SM
gauge group) with a mass $m_{\nu_s}$,  which only interact with the
active neutrinos and  
possibly the Higgs. Such sterile fermions 
can be simply added to the SM content, as is the case of 
right-handed (RH) neutrinos in type I seesaw
mechanisms of neutrino mass generation~\cite{seesaw:I}, 
or emerge in association
with extended gauge groups - as occurs in Left-Right (LR) symmetric
models~\cite{LR:orig}. 

Additional sterile fermions have been proposed at very different
scales, aiming at addressing very distinct observational problems:
very light states, with a mass around the eV, have long been
considered to explain the so-called ``oscillation anomalies''
(reactor, Gallium, and LSND); for a recent update, see~\cite{Gariazzo:2017fdh}. 
Sterile fermions with masses around the keV are natural warm dark
matter candidates. They are subject to stringent constraints,
concerning their stability, indirect detection (via X-ray emission),
phase space constraints, successful production in the early Universe,
and finally, impact for structure formation. Recent reviews of the
cosmological appeal of these states can be found 
in~\cite{Abazajian:2012ys,Adhikari:2016bei}.  
The MeV regime opens the first window to searches at the
high-intensity frontier: several observables are
already sensitive to sterile fermion masses around the MeV.
However, cosmological constraints remain severe, in particular those
arising from Big-Bang nucleosynthesis (BBN); the latter are
particularly stringent for $m_{\nu_s} \lesssim 200$~MeV. 

Heavy neutral leptons (HNL), with masses ranging from the GeV to the tenths
of TeV are among the phenomenologically most exciting extensions of
the SM, as they can give rise
to numerous phenomena. Sterile fermionic states with a mass
between 100 MeV and a few GeV can be produced in muon and tau decays,
as well as in meson leptonic and semileptonic decays, giving rise to
deviations from SM expectations, or to signatures which would
otherwise be forbidden in the SM, such as violation of lepton number,
of lepton flavour, or of  lepton universality.  These processes can 
be looked for in laboratory and high-intensity experiments.
Provided their mass is not
much larger than the electroweak (EW) scale, 
sterile fermions can also be present in
the decays of $Z$ and Higgs bosons, typically produced
in high-energy colliders, and thus induce new experimental
signatures. Finally, heavy sterile states can also be directly
produced at colliders ($m_{\nu_s}\sim$~a few TeV).
In all cases, they can contribute to numerous processes as
virtual intermediate states. 
In addition, sterile fermions in the GeV-TeV range
also open the door to explaining the BAU via low-scale scenarios of
leptogenesis (some relying on resonant mechanisms) without conflict
with other cosmological observations, such as BBN, for example.

Depending not only on their mass regime but also on their couplings to
the ``active'' neutrinos, the new neutral fermions can lead to very
distinctive phenomenological features, which in turn identify
the possible means to explore their presence:
additional neutral leptons can be
searched for in cosmology and astrophysics, in high energy colliders,
or in high-intensity experiments. 
Concerning the latter, the fermionic singlets can be responsible for
contributions to electric and magnetic leptonic dipole moments, and be
responsible for numerous rare transitions and decays, including
charged lepton flavour violating (cLFV), lepton number violating (LNV)
and lepton flavour universality violating (LFUV) observables. 

Observables involving muons offer numerous possibilities to look for imprints of
the HNL; this is the case of cLFV channels (rare decays and
transitions)~\cite{Kuno:1999jp,Bernstein:2013hba}, 
and contributions to leptonic moments (electric and
magnetic)~\cite{Raidal:2008jk}. 
The advent of very intense beams renders the muon system 
one of the best laboratories to look for NP states capable of 
contributions to the above mentioned rare decays and transitions.  

The phenomenological appeal of HNL 
is thus manifest: their non-negligible
contributions might either lead to ease some existing tensions between
SM predictions and experimental data, or then render these SM
extensions testable and even falsifiable. In what follows, we focus on
these NP candidates, and discuss the impact that they might have for
numerous observables which can be probed at the high-intensity
frontier. We will consider two complementary approaches, firstly
discussing the phenomenological impact of minimal SM extensions via
a number $n_S$ of heavy sterile states (bottom-up approach, or
``$3+n_S$ toy-models''), and subsequently consider contributions of
the HNL to several observables when the latter are naturally embedded
in the framework of complete models of New Physics.

This contribution is organised as follows: in
Section~\ref{sec:muon:high} we present the modified leptonic
interaction Lagrangian (due to the presence of the heavy neutral
leptons), and detail the contributions of the HNL to several
observables; we also briefly describe the constraints that these SM
extensions must comply with. Section~\ref{sec:thmodels} is devoted to
discussing the impact of these new sterile
fermionic states, first under a model-independent bottom-up approach,
and then for 
several well-motivated New Physics models embedding them. Our final comments and
discussion are collected in the Conclusions.

\section{Muon high-intensity observables}
\label{sec:muon:high}

As mentioned in the Introduction, the existence of heavy neutral
leptons is a well motivated hypothesis. The particular case of sterile
fermions - i.e., SM singlets which only interact with light active
neutrinos, other singlet-like states, and/or the Higgs sector 
- has received increasing
attention in recent years, due to the extensive impact they can have,
regarding both particle physics and cosmology.

\subsection{Impact of HNL on muon observables}\label{sec:muon.observables}
Depending on their masses and mixings with the 
light (active) neutrinos, 
sterile fermions have a potential impact on 
a number of high-intensity observables, in particular those involving
the muon sector; high-intensity muon beams may thus offer a unique
window to probe and to indirectly test  SM extensions via HNL.  

In order to address their phenomenological effects,
it is convenient to consider a modified SM Lagrangian, which reflects
the addition of $n_S$ sterile neutral fermions that mix with
the active neutrinos. In order to simplify this first approach, we
further hypothesise that:

\noindent
\quad (a) the new states are Majorana fermions;
 
\noindent
\quad (b) the interactions responsible for their mixing with the left-handed
(active) states lead to a generic mass term of the form
\begin{equation}\label{eq:steriles:th:Lmass}
\mathcal{L}_\text{mass}\, = \, 
\frac{1}{2}\, \nu^{\prime T}_{L} \, C^\dagger \, 
M_{\nu_L \nu_s}\, \nu^{\prime}_{L} \, + \text{H.c.}\,,
\end{equation}
which is written in the ``flavour basis'' (denoted with an ``$\prime$''
superscript); in the above, $C$ denotes the charge conjugation
matrix\footnote{We follow the conventions under which 
$C \gamma_\mu^T C^{-1} = -\gamma_\mu$, with $C^T = -C$. 
The fields transform as $\psi^c = C \bar \psi^T$, changing chirality
under the action of the operator, i.e., $\psi^c_R$ is a left-handed
field.}. 
The fields have been assigned as
\begin{equation}
\nu_L^\prime\, = \,
\left(
\nu^\ell_L,\, {\nu_R^{s}}^c 
\right)^T\,,
\quad \text{with} \quad 
\nu^\ell_L\, = \, 
\left(\nu_{e L},\,
\nu_{\mu L},\,
\nu_{\tau L}
\right)^T\, , \quad \text{and} \quad 
{\nu_R^{s}}^c\, = \, 
\left(
\nu^c_{s_1 R}\,,
\hdots\,,
\nu^c_{s_n R}\right)^T\, .
\end{equation}
In the above, $M_{\nu_L \nu_s}$ is a $(3+n_S)\times(3+n_S)$ matrix,
in general complex symmetric. 
The diagonalisation of the
latter allows to identify the $(3+n_S)$ physical (Majorana) neutrino fields, 
\begin{equation}\label{eq:steriles:th:UMU}
{U}_\nu^T \, M_{\nu_L \nu_s} \, {U}_\nu\, =\, 
\text{diag}(m_{\nu_1}, ..., m_{\nu_{3+n_S}})\, ,
\end{equation}
with the corresponding basis transformations, 
\begin{equation}\label{eq:steriles:th:nuUnu}
\nu_L^\prime\, = \, {U}_\nu\, \nu_L\,,
\quad \text{where} \quad
\nu_L^T\,=\,\left( \nu_{1L}, ..., \nu_{(3+n_S) L}\right)\,.
\end{equation}
In the physical basis, the Lagrangian mass term of
Eq.~(\ref{eq:steriles:th:Lmass}) can be rewritten as 
\begin{equation}
\mathcal{L}_\text{mass} \, = \, -\frac{1}{2} \sum_{k=1}^{3+n_S} 
m_k \, \bar \nu_k\, \nu_k\,, \quad 
\text{with}\ \ \ \nu_k = \nu_{k L} + \nu^c_{k L} \qquad
(\nu_k = \nu^c_k)\,,
\end{equation}
while the SM Lagrangian is modified
(in the Feynman-'t Hooft gauge)  
as follows\footnote{
See, for example,~\cite{Ilakovac:1994kj} for a detailed derivation 
starting from explicit lepton mass matrices.}: 
\begin{align}
& \mathcal{L}_{W^\pm}\, =\, -\frac{g_w}{\sqrt{2}} \, W^-_\mu \,
\sum_{\alpha=1}^{3} \sum_{j=1}^{3 + n_S} {\bf U}_{\alpha j} \bar \ell_\alpha 
\gamma^\mu P_L \nu_j \, + \, \text{H.c.}\,, \label{eq:lagrangian:W}\\
& \mathcal{L}_{Z^0}\, = \,-\frac{g_w}{4 \cos \theta_w} \, Z_\mu \,
\sum_{i,j=1}^{3 + n_S} \bar \nu_i \gamma ^\mu \left(
P_L {\bf C}_{ij} - P_R {\bf C}_{ij}^* \right) \nu_j\,, \label{eq:lagrangian:Z} \\
& \mathcal{L}_{H^0}\, = \, -\frac{g_w}{2 M_W} \, H  \,
\sum_{i,j=1}^{3 + n_S}  {\bf C}_{ij}  \bar \nu_i\left(
P_R m_i + P_L m_j \right) \nu_j + \, \text{H.c.}\,  \label{eq:lagrangian:H}\\
& \mathcal{L}_{G^0}\, =\,\frac{i g_w}{2 M_W} \, G^0 \,
\sum_{i,j=1}^{3 + n_S} {\bf C}_{ij}  \bar \nu_i  
\left(P_R m_j  - P_L m_i  \right) \nu_j\,+ \, \text{H.c.},  \nonumber  \\
& \mathcal{L}_{G^\pm}\, =\, -\frac{g_w}{\sqrt{2} M_W} \, G^- \,
\sum_{\alpha =1}^{3}\sum_{j=1}^{3 + n_S} {\bf U}_{\alpha j}   \bar \ell_\alpha \left(
m_{\ell_\alpha} P_L - m_j P_R \right) \nu_j\, + 
\, \text{H.c.}\,. \label{eq:lagrangian:G}
\end{align}
in which 
$P_{L,R} = (1 \mp \gamma_5)/2$, $g_w$ and 
$\theta_w$ respectively denote the weak coupling constant and weak
mixing angle,
and $m_j$ are the physical neutrino masses ($j=1,...,3 + n_S$). 
We have also introduced  
${\bf C}_{ij}=\sum_{\alpha=1}^{3} {\bf U}_{\alpha i}^*{\bf U}_{\alpha
  j}$, where $ {\bf U}$ 
is a $3\times (3+n_S)$ (rectangular) matrix, which can be decomposed as
\begin{equation}\label{eq:U:eta:PMNS2}
{\bf U} \,= \,\left(\tilde U_\text{PMNS} \, , \, U_{\nu S} \right) \, .
\end{equation}
In the above, $\tilde U_\text{PMNS}$ is a $3 \times 3$ matrix and $U_{\nu S}$ is
a $3 \times (n_s)$ matrix. 
The matrix $U_{\nu S}$
encodes the information about the mixing between 
the active neutrinos and the sterile singlet states (which can be
often approximated, in particular in type I seesaw-like models, as 
$U_{\nu S} \approx \sqrt{m_\nu/m_N}$); the 
left-handed mixings are parametrised by a non-unitary $\tilde
U_\text{PMNS}$  introduced in Eq.~(\ref{eq:U:eta:PMNS2}), 
which can be cast
as~\cite{FernandezMartinez:2007ms}  
\begin{equation}
\tilde U_\text{PMNS} \, = \,(\mathbb{1} - \eta)\, U_\text{PMNS}\,,
\nonumber
\end{equation}
where the  matrix $\eta$ encodes the deviation of $\tilde
U_\text{PMNS}$ from unitarity~\cite{Schechter:1980gr,Gronau:1984ct}, 
due to the presence of extra fermion states.
In the limiting case of three neutrino generations (the 3 light active 
neutrinos), and 
assuming alignment of the charged lepton's weak and 
mass bases, ${\bf U}$ can be identified with the 
(unitary) PMNS matrix, $U_\text{PMNS}$.

In summary, the presence of the additional states leads to 
the violation of lepton flavour in both charged and neutral current
interactions. The above modified interactions are at the source of new
contributions to many observables, which we proceed to discuss.

\subsubsection{Lepton dipole moments: muon EDM and 
$\pmb{(g-2)_\mu}$}\label{sec:edm.g2}

Should the model of NP involving heavy sterile fermions further
include sources of CP violation, then one expects that there will be
non-negligible contributions to electric dipole moments (EDMs), 
which violate both T and CP conservation. 
Likewise, one also expects new contributions to flavour
conserving observables - as for example the anomalous magnetic moment of
the muon.

\paragraph{Electric dipole moments}

The current bound for the muon EDM is 
$|d_\mu| / e \lesssim 1.9 \times 10^{-19}$~cm (Muon
$g-2$~\cite{Bennett:2008dy}), and the future expected sensitivity 
should improve to $\mathcal{O}(10^{-21})$~cm  (J-PARC
$g-2$/EDM~\cite{Saito:2012zz}).
In general, the contributions of heavy leptons to the EDMs 
occur at the two-loop level, and call upon a minimal content of at least 2
non-degenerate sterile states~\cite{Abada:2015trh}. 
As shown in~\cite{Abada:2015trh}, in the presence of $n_S$ new states,
the EDM of a charged lepton $\ell_\alpha$ can be written as
\begin{equation}
d_\alpha\,=\,-\frac{g_2^4\ e\ m_\alpha}{4\,(4\pi)^4\,M_W^2}
\sum_\beta\sum_{i,j} \left[
J_{ij\alpha\beta}^{M}\,I_M\left(x_i,x_j,x_\alpha,x_\beta\right) 
\,+\,J_{ij\alpha\beta}^{D} \,I_D\left(x_i,x_j,x_\alpha,x_\beta\right)
\right],
\label{eq:edm.lepton}
\end{equation}
in which $e$ is the electric charge, 
$g_2$ is the SU(2) coupling constant, and $m_\alpha$ ($M_W$) 
denote the mass of the charged lepton ($W$ boson mass). In the above
$J_{ij\alpha\beta}^{M,D}$ are invariant quantities - respectively
sensitive to Majorana and Dirac CP violating phases, defined as  
\begin{equation}
J_{ij\alpha\beta}^M \,=\,\text{Im}\left(U_{\alpha j}U_{\beta j}U_{\beta
i}^*U_{\alpha i}^*\right)\ \quad \text{and}\quad 
J_{ij\alpha\beta}^D\,=\,\text{Im}\left(U_{\alpha j}U_{\beta j}^{*}U_{\beta
i}U_{\alpha i}^*\right)\,, 
\label{eq:edm:JUMD}
\end{equation}
and $I_{M,D}$ are the loop
functions cast in terms of $x_A\equiv m_A^2/m_W^2$
($A=i,j,\alpha,\beta$) (see~\cite{Abada:2015trh}).  
As will be illustrated via the phenomenological analyses summarised in 
Section~\ref{sec:thmodels}, the ``Majorana''-type contributions tend
to dominate over the ``Dirac''ones.

\paragraph{Anomalous magnetic moments}

The muon anomalous magnetic moment induced at
one-loop level by neutrinos and the $W$ gauge boson  is 
\begin{equation}\label{eq:g-2}
a_\mu\,=\,\frac{\sqrt{2}\,G_F\,m_\mu^2}{\left(4\,\pi\right)^2}
\sum_{i=1}^{3+n_S}\left|U_{\mu i}\right|^2\,
F_M\left(\frac{m_i^2}{M_W^2}\right)\,,
\end{equation}
in which $G_F$ is the Fermi
constant, $m_\mu$ the muon mass, and $m_{i}$ refers to the mass of the
neutrinos in the loop; the 
loop function $F_M(x)$ is defined in the
Appendix~\ref{appendix.loop}, Eq.~(\ref{eq:FM}).  
Subtracting the SM-like contribution from the full expression of 
Eq.~(\ref{eq:g-2}) (arising from one-loop diagrams, dominated by
the mostly active light neutrino contribution), one obtains 
\begin{equation}
\Delta{a}_\mu\,\approx\,
-\frac{4\,\sqrt{2}\,G_F\,m_\mu^2}{(4\pi)^2}\,
\sum_{i=4}^{3+n_S}|U_{\mu\,i}|^2G_\gamma\left(\frac{m_i^2}{M_W^2}\right),
\label{eq:mag}
\end{equation}
where one neglects the light neutrino masses $m_i~(i=1,2,3)$ and
$G_\gamma(x)$ is also given in Appendix~\ref{appendix.loop},
Eq.~(\ref{eq:steriles:Gloop}). 
The 
experimental value of the muon anomalous magnetic moment has
been obtained by the Muon $g-2$ Collaboration~\cite{Bennett:2006fi}, and
the discrepancy between the experimental value and the SM prediction is given
by~\cite{Patrignani:2016xqp}
$ \Delta{a}_\mu\equiv a_\mu^\mathrm{exp}-
a_\mu^\mathrm{SM}=2.88\times10^{-9}$.
As shown in~\cite{Abada:2014nwa}, 
the new contributions from HNL to the muon anomalous magnetic moment 
cannot account for the discrepancy between the experimental measured
value and the SM theoretical prediction.

\subsubsection{Charged lepton flavour violation: the muon sector}
\begin{figure}
\begin{center}
\begin{tabular}{ccc}
\includegraphics[width=0.33\textwidth]{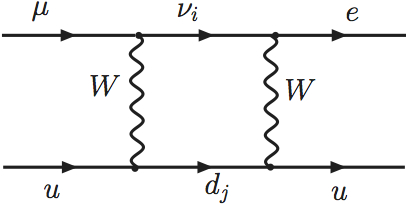} & 
\includegraphics[width=0.27\textwidth]{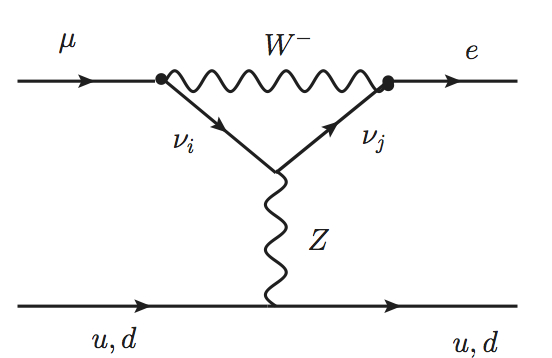} &
\raisebox{-2mm}{\includegraphics[width=0.35\textwidth]{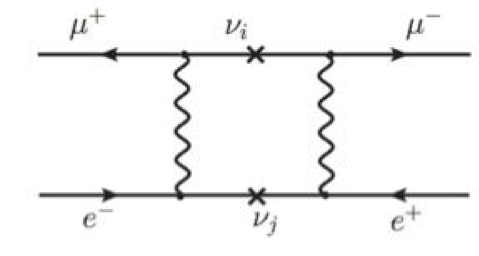}}
\end{tabular}
\end{center}
\caption{Examples (subset) of HNL mediated 
diagrams contributing to some of the cLFV decays and transitions 
discussed in the text: 
nuclear $\mu-e$ conversion and muonium oscillations. 
In the neutral fermion internal lines, $i,j=1,..., 3+n_S$.}  
\label{fig:sterile:muediagrams}
\end{figure}

Many new contributions to cLFV rare decays and transitions involving
muons can be induced by the modified neutral and charged lepton currents;
examples of Feynmann diagrams mediated by HNL at the origin of the
cLFV transitions can be found in Fig.~\ref{fig:sterile:muediagrams}.
On Table~\ref{tab:cLFV-muon} we
summarise the experimental status (current bounds and future
sensitivities) of several processes involving muons which can be
studied at high-intensity frontier.

{\small
\renewcommand{\arraystretch}{1.1}
\begin{table}[h!]
\begin{center}
\begin{tabular}{|c|c|c|}
\hline
cLFV process & Current bound & Future sensitivity   \\	
\hline
\hline
$\text{BR}(\mu^+\to e^+ \gamma)$	&  
$ 4.2\times 10^{-13}$ (MEG~\cite{TheMEG:2016wtm}) 	&  
$6\times 10^{-14}$ (MEG II~\cite{Baldini:2018nnn})  	\\
\hline
$\text{BR}(\mu^+ \to e^+ e^- e^+)$	&  
$1.0\times 10^{-12}$ (SINDRUM~\cite{Bellgardt:1987du}) 	& 
$10^{-15(16)}$ (Mu3e~\cite{Blondel:2013ia})  	\\
\hline
$\text{CR}(\mu^-- e^-, \text{N})$ & 
$ 7 \times 10^{-13}$ (Au, SINDRUM~\cite{Bertl:2006up}) & 
$10^{-14}$ (SiC, DeeMe~\cite{Nguyen:2015vkk})    \\
& & $10^{-15 (-17)}$ (Al, COMET~\cite{Kuno:2013mha,Krikler:2015msn})  \\
& & $3 \times 10^{-17}$ (Al, Mu2e~\cite{Bartoszek:2014mya}) \\
& & $10^{-18}$ (Ti, PRISM/PRIME~\cite{Kuno:2005mm})  \\
\hline
\begin{tabular}{l}
$\text{CR}(\mu^- +  \text{Ti}\to e^+ +  \text{Ca}^*)$\\
$\text{CR}(\mu^- +  \text{Ti}\to e^+ +  \text{Ca})$ 
\end{tabular}
& 
\begin{tabular}{l}
$ 3.6 \times 10^{-11}$ (SINDRUM~\cite{Kaulard:1998rb}) \\ 
$ 1.7 \times 10^{-12}$ (SINDRUM~\cite{Kaulard:1998rb})
\end{tabular}
& --
\\
\hline
$\text{P}(\text{Mu}-\overline{\text{Mu}})$ & $8.3 \times
10^{-11}$ (PSI~\cite{Willmann:1998gd}) &  -- \\ 
\hline
\end{tabular}
\caption{Current experimental bounds and 
future sensitivities of cLFV processes relying on intense muon beams.}
\label{tab:cLFV-muon}
\end{center}
\end{table}
\renewcommand{\arraystretch}{1.}
}

\paragraph{Muon radiative decays: $\pmb{\mu \to e \gamma}$}
In a framework with a total number of $3+n_S$ physical neutral
leptons, the contributions to the cLFV radiative decays 
$\ell_i \to \ell_j \gamma$
can be written as 
\begin{equation}\label{eq:steriles:BR.liljgamma}
\text{BR}(\ell_i \to \ell_j \gamma)\, =\, 
\frac{\alpha_w^3 \, \sin \theta_w}{256 \, \pi^2}\, 
\frac{m_{\ell_i}^4}{M_W^4}\, 
\frac{m_{\ell_i}}{\Gamma_{\ell_i}}\,
\left| G_\gamma^{\ell_i \ell_j}(x_{k})\right|^2\,,
\end{equation}
with  $\alpha_w=g_w^2/(4\pi)$,
($s_w$ corresponding to the sine of the weak
mixing angle), and where $m_{\ell_i}$ and $\Gamma_{\ell_i}$ 
denote the mass and decay
width of the decaying lepton. For the case of the muon, the latter is
given by~\cite{Beg:1982ex} 
\begin{equation}\label{eq:muon:decay:width}
\Gamma_\mu\, =\, \frac{G^2_F\, m_\mu^5}{192 \pi^3}\,
\left(1-8 \frac{m_e^2}{m_\mu^2}\right) \, \left[
1 + \frac{\alpha_\text{em}}{2 \pi} \left( \frac{25}{4} - \pi^2
\right) \right]\,,
\end{equation}
In Eq.~(\ref{eq:steriles:BR.liljgamma}),  
$G_\gamma^{\ell_i \ell_j}(x_{k})$ denotes a composite form factor
which encodes the lepton mixing angles and which is given in 
Appendix~\ref{appendix.form},
while the corresponding loop function, written in terms of 
 $x_{k} = m^2_{\nu_k}/M^2_W$, can be found 
in Appendix~\ref{appendix.loop} (see Eq.~(\ref{eq:steriles:Gloop})).
The limits for the form factors, as well as for the different loop
functions, which apply in extreme regimes (e.g. $x \gg 1$, or strong
hierarchy in the sterile spectrum) can be found in the pioneering study
of~\cite{Ilakovac:1994kj}.

\paragraph{Muon 3-body decays: $\pmb{\mu \to 3e}$}
The full formulae, detailing the most general
decay $\ell_i \to \ell_j \ell_k \ell_m$ in the presence of sterile
states can be found
in~\cite{Ilakovac:1994kj}; here we mostly focus on cases 
with same-flavour final
states. The branching ratio for the decay 
$\ell_i \to 3 \ell_j$ is given by
\begin{align}
\text{BR}(\ell_i \to 3 \ell_j) \, &=\, 
\frac{\alpha_w^4}{24576 \pi^3} \,\frac{m_{\ell_i}^4}{M_W^4}\,
\frac{m_{\ell_i}}{\Gamma_{\ell_i}}\, \left\{ 2 \left| 
\frac{1}{2} F_\text{box}^\text{3body} + F_Z^\text{3body} 
- 2 \sin^2 \theta_w (F_Z^\text{3body} - F_\gamma^\text{3body})^2 
\right|^2 +\right. \nonumber \\
&+ 4 \sin^2 \theta_w |F_Z^\text{3body} - F_\gamma^\text{3body}|^2
+16\sin^2 \theta_w \, \text{Re}\left[\left(
F_Z^\text{3body} + \frac{1}{2} F_\text{box}^\text{3body}\right)\,
G_\gamma^{\text{3body}*} \right] - \nonumber \\
& \left.
-148 \sin^2 \theta_w \, \text{Re}\left[\left(
F_Z^\text{3body} -F_\gamma^\text{3body}\right)\,
G_\gamma^{\text{3body}*} \right] + 
32 \sin^2 \theta_w \,|G_\gamma^{\text{3body}}|^2\, 
\left[ \ln \frac{m_{\ell_i}^2}{m_{\ell_j}^2} -\frac{11}{4}
\right] \right\}\,,
\end{align}
where one has included the contributions from (non-local) dipole, photon and $Z$
penguins as well as box diagrams, corresponding to the composite form
factors $G_\gamma^{\text{3body}}$, 
$F_\gamma^\text{3body}$, $F_Z^\text{3body}$ and
$F_\text{box}^\text{3body}$ (see Appendix~\ref{appendix.form}).

\paragraph{Neutrinoless muon-electron conversion in nuclei}
Muonic atoms are formed when a negatively charged muon is stopped
inside matter, and after cascading down in energy level becomes bound
in the 1s state; in the presence of NP, the
muon can be converted into an electron without neutrino emission
(neutrinoless muon capture, or conversion). 
The observable can be defined as 
\begin{equation}\label{eq:CR:def}
\text{CR}(\mu -e, \text{ N}) \, = \frac{\Gamma (\mu^- + N \to e^- 
+N)}{\Gamma (\mu^- + N \to \text{ all captures)}}\,;
\end{equation}
the rate of the 
coherent conversion (spin-independent process)\footnote{For a
  recent study of spin-dependent contributions to muon-electron
  conversion in nuclei, see~\cite{Cirigliano:2017azj}.} 
increases with the atomic number ($Z$) for nuclei with  
$Z\lesssim 30$, being maximal for $30 \lesssim Z \lesssim
60$~\cite{Kitano:2002mt}.
For heavier elements, one finds a
reduction of the corresponding conversion rate (due to 
Coulomb distortion effects of the wave function).

In the framework of the 
SM extended by sterile neutrinos, the contributions to the
muon-electron conversion rate can be written as (see, 
for example,~\cite{Alonso:2012ji,Abada:2015oba})
\begin{equation}
\text{CR}(\mu-e, \text{N}) =
\frac{2\,G_F^2\,\alpha_{w}^2\,m_\mu^5}{(4\pi)^2
  \,\Gamma_\text{capt}(Z)}
\left|4\,V^{(p)}\left(2 \,\tilde{F}_{u}^{\mu e}+\tilde{F}_{d}^{\mu
  e}\right)+
4\,V^{(n)}\left(\tilde{F}_u^{\mu e}+2\,\tilde{F}_{d}^{\mu e}\right)
+ D \,G^{\mu e}_{\gamma} \frac{s^2_w}{2 \sqrt{4 \pi \alpha}}  \right|^2\,,
\label{eq:CRmue}
\end{equation}
in which $\alpha=e^2/(4\pi)$ and 
$\Gamma_\text{capt}(Z)$ is the capture rate of the
nucleus (with an atomic number $Z$)~\cite{Kitano:2002mt},  
and the form factors $\tilde{F}_{q}^{\mu e}$ ($q=u,d$) are given by 
\begin{equation}
\tilde F_q^{\mu e}\, =\, 
Q_q \, s_w^2 F^{\mu e}_\gamma+F^{\mu e}_Z
\left(\frac{{I}^3_q}{2}-Q_q\, s_w^2\right)+
\frac{1}{4}F^{\mu eqq}_\text{Box}\,,
\label{eq:tildeFqmue}
\end{equation}
where $Q_q$ corresponds to
the quark electric charge ($Q_u=2/3, Q_d=-1/3$) and
${I}^3_q$ is the weak isospin (${I}^3_u=1/2\,,\,
{I}^3_d=-1/2$). 
The form factors $D$, $V^{(p)}$ and $V^{(n)}$ encode the 
relevant nuclear information. 
The quantities $F^{\mu e}_\gamma$, $F^{\mu e}_Z$ and $F^{\mu
  eqq}_\text{Box}$ denote the form factors of the distinct 
diagrams contributing to the process (see examples in 
Fig.~\ref{fig:sterile:muediagrams}), their expressions 
 being given in Eqs.~(\ref{eq:FF}-\ref{Fmueee}) of 
Appendix~\ref{appendix.form}, and those of the
involved loop factors can also be found in 
Appendix~\ref{appendix.loop}.

\paragraph{LNV muon-electron conversion: $\pmb{\mu^- - e^+}$, N}

Should the heavy leptons be of Majorana nature, then they can induce a
cLFV and LNV conversion process in the presence of nuclei, 
$\mu^- + (A,Z) \rightarrow e^+ + (A,Z-2)^*$. 
Contrary to the lepton number conserving process, in this case the 
final state nucleus can be in a state different from the initial
one (in particular, it can be either in its ground state
or in an excited one - thus preventing a coherent enhancement).
We will not address this process here (for
model-independent recent approaches,
see~\cite{Geib:2016atx,Berryman:2016slh}).

\paragraph{Coulomb enhanced muonic atom decay: $\pmb{\mu^- e^- \to e^-
  e^-}$}

In the presence of NP, another cLFV channel can be studied for muonic atoms:
their Coulomb enhanced decay into a pair of
electrons~\cite{Koike:2010xr,Uesaka:2017yin},  
\begin{equation}\label{eq:me2ee} 
\mu^-\, +\, e^-\, \to \, e^-\, e^-\,,
\end{equation}
in which the initial fermions are the muon and the atomic 1s electron,
bound in the Coulomb field of the nucleus. 
(This is a ``new'' observable, which has not yet been experimentally
searched for; thus no experimental bounds are currently available.)
In SM extensions via heavy neutral fermions, the dominant
contributions arise from contact interactions, 
which include photon- and $Z$-penguins as well as box
diagrams\footnote{In this class of models, and in 
  the regimes  associated with significant cLFV
  contributions, the ``long-range'' photonic interactions are typically
subdominant.}. 
Neglecting the
interference between contact terms (which can be sensitive to CP
violating phases), the new contributions of the sterile states 
to the cLFV decay of a muonic atom, with an atomic number $Z$, can be
written as
\begin{align}
&\text{BR}(\mu^- e^- \to e^- e^- , \text{ N})\, \equiv 
\, \tilde{\tau}_{\mu}\,
  \Gamma(\mu^- e^- \to e^- e^- , \text {N}) \nonumber \\ 
& = 24\pi \, f_\text{Coul.}(Z)\, 
\alpha_w  \left( \frac{m_e}{m_{\mu}} \right)^{3}\,
  \frac{\tilde{\tau}_{\mu}}{\tau_{\mu}}\,
  \left(16 \, \left | \frac{1}{2} \left (\frac{g_w}{4 \pi} \right)^2
  \left (\frac{1}{2}F^{\mu eee}_\text{Box} + F_Z^{\mu e} - 
  2 \sin^2\theta_w \left (F_Z^{\mu e} - F_\gamma^{\mu e} \right)
  \right) \right|^2 + \right . \nonumber\\ 
& \left . \phantom{+}  
4  \, \left| \frac{1}{2} \left(\frac{g_w}{4 \pi} \right)^2 
  2 \sin^2\theta_w \left (F_Z^{\mu e} - F_\gamma^{\mu e} \right)
 \right|^2 \right) \, .
\label{eq:br-4fermi}
\end{align}
In the above, 
$F_{\gamma,Z}^{\mu e}$  
correspond to the contributions from photon- and $Z$-penguins
(as previously introduced in Eq.~(\ref{eq:CRmue})); 
$\tilde{\tau}_{\mu}$ denotes the lifetime of the muonic atom,
that depends on the specific element from which it is formed
(always smaller than the lifetime of free muons, $\tau_{\mu}$). 
The function $f_\text{Coul.}(Z)$ encodes the effects of the
enhancement due to the Coulomb attraction from the nucleus (which
increases the overlap of the 1s electron and muon wavefunctions);
typically, $f_\text{Coul.}(Z) \propto (Z-1)^3$, or even more than
$(Z-1)^3$ for large $Z$ nuclei~\cite{Uesaka:2017yin}.

\paragraph{Muonium channels: $\pmb{\text{Mu}-\overline{\text{Mu}}}$
  oscillation and $\pmb{\text{Mu} \to e^+ e^-}$ decay}

The Muonium (Mu) atom is a Coulomb  
bound state of an electron and an anti-muon
($e^-\mu^+$)~\cite{Pontecorvo:1957cp}; strongly resembling an
hydrogen-like atom, 
its binding is purely electromagnetic,
and thus can be well described by SM electroweak interactions, 
with the advantage of being
free of hadronic uncertainties. 
The Muonium system is thus an interesting laboratory to 
test for 
the presence of new states and modified interactions.
Concerning cLFV, two interesting channels can be studied:
Muonium-antimuonium conversion 
Mu-$\overline{\text{Mu}}$~\cite{Feinberg:1961zza}, and the muonium's 
decay to an electron-positron pair, 
$\text{Mu}\to e^+ e^-$.

Under the assumption of $(V -A) \times (V - A)$ interactions, the 
Mu-$\rm \overline{Mu}$ transition can be described by an effective
four-fermion interaction with a coupling constant $G_{\rm M\overline{M}}$, 
\begin{equation}
 \mathcal{L}_\text{eff}^{\rm M\overline{M}} \,= \, \frac{G_{\rm
     M\overline{M}}}{ \sqrt{2} } 
\left[\, {\overline \mu}\,  \gamma^{\alpha} (1 - \gamma_5) \,e
\,\right] \left[ \,{\overline \mu}\, \gamma_{\alpha} (1 - \gamma_5)\,
e \,\right] \, .
\label{eq:Leff_muonium}
\end{equation}
Searches for Mu-$\rm \overline{Mu}$ conversion at PSI have allowed to
establish the current best bound on 
$G_{\rm M\overline{M}}$~\cite{Willmann:1998gd}: 
$\left |\text{Re}\left( G_{\rm M\overline{M}} \right) \right| \leq 3.0
\times 10^{-3}G_F$, at 90\%C.L.~\cite{Willmann:1998gd}. 
In SM extensions including HNL,  Mu-$\rm \overline{Mu}$ conversion
receives contributions from four distinct types of box diagrams
(mediated by Dirac and Majorana neutrinos, see~\cite{Abada:2015oba}).
In a unitary gauge, the computation of these diagrams leads to
  the following expression for the  
effective coupling 
$G_{\rm M\overline{M}}$~\cite{Clark:2003tv,Cvetic:2005gx,Liu:2008it,Abada:2015oba}:
\begin{equation}
\frac{G_{\rm M\overline{M}}}{\sqrt{2}}\,=\,
-\frac{G_F^2 M_W^2}{16\pi^2}\left[\sum_{i,j =1}^{3+n_S}
({\bf U}_{\mu i}\,{\bf U}^\dagger_{e i})\,({\bf U}_{\mu j}\,
{\bf U}^\dagger_{e j})\,{\rm G}_{\rm Muonium}(x_i, x_j)\right]\,,
\label{eq:Gmumu}
\end{equation}
in which $x_i =\frac{m_{\nu_i}^2}{M_W^2}, i=1,...,3+n_S$ and ${\rm
  G}_{\rm Muonium}(x_i, x_j)$ is the loop 
function arising from the two groups of boxes (generic
and Majorana), and is given in Appendix~\ref{appendix.loop}. 

The presence of HNL can also be at the origin of the cLFV Muonium
decays~\cite{Cvetic:2006yg,Abada:2015oba};
the decay ratio can be written as 
\begin{equation}\label{eq:Mudecay:BR}
\text{BR}(\text{Mu} \to e^+ e^-) \, =\, 
\frac{\alpha_\text{em}^3}{\Gamma_\mu \, 32 \pi^2}\, 
\frac{m_e^2 m_\mu^2}{(m_e + m_\mu)^3}\,
\sqrt{1 -4\, \frac{m_e^2}{(m_e + m_\mu)^2}}\,
|\mathcal{M}_\text{tot}|^2\, ,
\end{equation}
with $\Gamma_\mu$ the muon decay width,
and where $|\mathcal{M}_\text{tot}|$ denotes the full amplitude, summed
(averaged) over final (initial) spins~\cite{Cvetic:2006yg}.
The full expression for $|\mathcal{M}_\text{tot}|$ can be found
in~\cite{Abada:2015oba}.  
At present, no bounds exist on this cLFV observable, nor are there
prospects for searches in the near future.

\paragraph{In-flight (on-target) conversion: 
$\pmb{\mu \to \tau}$ }

The advent of high-intensity and sufficiently energetic 
muon beams (for instance at muon and future neutrino factories) allows 
the study of 
another muon cLFV observable: in-flight (elastic) 
conversion of muons to taus, $\mu + N \to \tau +N$
(with $N$ denoting a generic nucleus)~\cite{Gninenko:2001id}. 
The $\ell_i \to \ell_j$ on-target conversion can be mediated (for
example) by photon and $Z$ boson exchanges; the differential cross sections
for $\gamma$-dominated and $Z$-only mediation can be respectively cast
as 
\begin{equation}\label{eq:photon:dsigma.dQ2:photon}
\left. \frac{d \sigma^{i\to j}}{d Q^2} \right|_{\gamma}\, =\, 
\frac{\pi \, Z^2\, \alpha^2}{Q^4 \, E_\text{beam}^2}\, 
H^\gamma_{\mu \nu}\, L^{\gamma \mu \nu}_{ij}\,,
\quad 
\left. \frac{d \sigma^{i\to j}}{d Q^2} \right|_{Z}\, =\, 
\frac{G_F^2}{32 \,\pi\, E_\text{beam}^2}\, 
H^Z_{\mu \nu}\, L^{Z \mu \nu}_{ij}\,,
\end{equation} 
in which $Q^2$ is the momentum transfer, 
$Z$ denotes the target atomic number,  $H^{\gamma,Z}_{\mu
  \nu}$ denotes the hadronic tensor and the 
leptonic tensors can be decomposed as 
$L^{\gamma (Z) \mu \nu}_{ij}=\, L^{\gamma (Z)}_{ij} L^{\gamma(Z) \mu
  \nu}(k, q)$, with 
$L^{\gamma (Z)}_{ij}$ encoding the cLFV (effective) couplings
(for a complete discussion and detailed list of contributing diagrams,
see~\cite{Abada:2016vzu}). 
In the presence of new heavy sterile fermions
$L^\gamma_{ij}$ and $L^Z_{ij}$ can be cast (we consider the case when
the target is made of nucleons) as  
\begin{equation}\label{eq:photon:dsigma.dQ2:photon:Lij}
L^\gamma_{ij}\, =\, \frac{\alpha_w^3\, s_w^2}{64\, \pi \,e^2}\,
\frac{m_{\ell_j}^2}{M_W^4}\, \left| G^\gamma_{ji}\right|^2\,,
\quad 
L^Z_{ij}\, =\, \frac{\alpha_w^4}{G_F^2\, M_W^4} 
\frac{2(-1/2 + \sin^2_w)^2 + \sin^4_w}{64}\, \left| F^Z_{ji}\right|^2\, ,
\end{equation}
with the associated cLFV form factors already having been introduced
for other observables.

\subsection{The several constraints on HNL}\label{sec:constraints}
The impact of the additional neutral leptons concerns not only
potentially observable contributions to 
the muonic processes discussed 
above, but also to several other observables, 
possibly in conflict with current data. It is thus mandatory to
evaluate the impact of these SM extensions in what concerns many 
available constraints obtained 
from high-intensity, high-energy, as well as from cosmology.

In addition to complying with neutrino data, 
i.e. mass differences and bounds on the PMNS mixing
matrix~\cite{Gonzalez-Garcia:2015qrr,Esteban:2016qun}, sterile
fermions can induce important contributions to  several EW observables due
to the modification of the charged and neutral currents.  
Other than respecting the perturbative unitarity 
condition~\cite{Chanowitz:1978mv,Durand:1989zs,Korner:1992an,
Bernabeu:1993up,Fajfer:1998px,Ilakovac:1999md}, 
$\frac{\Gamma_{\nu_i}}{m_{\nu_i}}\, < \, \frac{1}{2}\, 
(i=1,3+n_S)$,\footnote{Since the dominant contribution to
  ${\Gamma_{\nu_i}}$ arises from the charged current term, 
  one can rewrite the perturbative unitarity condition as:
$m_{\nu_i}^2\,\sum_{\alpha=1}^{3} {\bf U}_{\alpha i}^*\,
{\bf U}_{\alpha i}\,< 8 \pi \,{M^2_W}/{g^2_w}\, \ \ 
(i \geq 4)$,
with ${\bf U}$ the lepton mixing matrix. } 
bounds from electroweak precision tests~\cite{Akhmedov:2013hec,Basso:2013jka,
Fernandez-Martinez:2015hxa,Abada:2013aba} and 
non-standard
interactions~\cite{Antusch:2008tz,Antusch:2014woa,Blennow:2016jkn}
must also be taken into account.

The so-far negative searches for rare cLFV lepton decays and
transitions (among which those
discussed in the previous section), already put severe constraints on
SM extensions with additional
HNL~\cite{Ma:1979px,Gronau:1984ct,Ilakovac:1994kj,Deppisch:2004fa,Deppisch:2005zm,Dinh:2012bp,Alonso:2012ji,Abada:2014kba,Abada:2015oba};
likewise, at higher energies, searches for cLFV Higgs
decays~\cite{Arganda:2014dta,Deppisch:2015qwa,Banerjee:2015gca,BhupalDev:2012zg,Cely:2012bz,Bandyopadhyay:2012px}
and for neutral $Z$ boson decays~\cite{Illana:2000ic,Abada:2014cca,
  Abada:2015zea,DeRomeri:2016gum} give rise to further constraints,
which must then be taken into account. 
Several observables associated with leptonic and semi-leptonic meson decays  
(cLFV, LNV and LFUV) are also sensitive to new contributions from
sterile fermions, and the corresponding bounds must thus be taken into
account~\cite{Shrock:1980vy,Shrock:1980ct,Atre:2009rg,Abada:2012mc,Abada:2013aba,Abada:2017jjx}. 
Finally, and as discussed previously, there might be non-negligible
contributions from Majorana HNL to
CP violating observables, 
such as charged lepton
EDMs~\cite{deGouvea:2005jj,Abada:2014nwa,Abada:2015trh}, and to
neutrinoless double beta decays 
($0\nu 2 \beta$)~\cite{Deppisch:2012nb}\footnote{
Working in the framework of the SM extended by $n_s$ sterile fermions, 
one must generalise the definition of the effective mass (to which
the $0\nu 2 \beta$ amplitude is proportional to) as
$m_{ee}\, = 
\sum_{i=1}^{3+n_s} \frac{{\bf U}_{e i}\, m_i\, {\bf U}_{e i}}{1 -
  m_i^2/p^2\, + i\, m_i \Gamma_i/p^2}$, where $p^2 \simeq - 
(125 \mbox{ MeV})^2$ is the virtual momentum of the propagating
neutrino (obtained from average estimates over different decaying
nuclei)~\cite{Blennow:2010th}. The new mixings (and the possibility of
additional CP-violating Majorana phases) can have a sizeable impact on 
the effective mass.}.
The additional mixings and possible
new CP-violating Majorana phases might enhance the effective mass, potentially
rendering it within experimental reach, or even in conflict current bounds.
Further constraints arise from peak searches in meson 
decays~\cite{Shrock:1980vy,Shrock:1980ct,Lello:2012gi,Lazzeroni:2017fza,CortinaGil:2017mqf};
one should also apply the bounds arising from negative
searches for monochromatic lines in the
spectrum of muons from  $\pi^\pm \to \mu^\pm
\nu$~\cite{Kusenko:2009up,Atre:2009rg}, 
as well as those from direct searches at the LHC. 

Finally, HNL are also subject to constraints of cosmological origin:
a wide variety of cosmological
observations~\cite{Smirnov:2006bu,Kusenko:2009up} has been shown to
lead to severe bounds on heavy neutral leptons with a mass below the TeV
(obtained under the assumption of a standard cosmology). 
Mixings between the active neutrinos and the sterile fermions can lead
to radiative decays $\nu_i \to \nu_j
\gamma$, well constrained by cosmic X-ray searches;  
Large Scale Structure and Lyman-$\alpha$ data further constrain the
HNL states, since these can 
constitute a non-negligible fraction of the dark
matter of the Universe (thus impacting structure formation). 
Further bounds on the HNL masses and mixings with the active states
can be inferred from Lyman-$\alpha$
limits, the existence of additional degrees of freedom at the epoch of
Big Bang Nucleosynthesis, and Cosmic Microwave Background data (among
others). 
Notice however, that in scenarios of ``non-standard cosmology''
(for instance, in the case of 
low reheating temperatures~\cite{Gelmini:2008fq}, or when the heavy
neutral leptons couple to  a dark sector~\cite{Dasgupta:2013zpn}), 
all the above cosmological bounds can be evaded.

\section{Phenomenological implications of HNL for muon 
observables}\label{sec:thmodels}
Sterile neutrinos are well-motivated New Physics candidates, and their
existence is considered at very different mass scales, as motivated 
by distinct observations.  As mentioned before, 
heavier states, with a mass ranging from the MeV to  a few TeV,  are
particularly appealing, as  
they can give rise to numerous phenomena which can be looked for in
laboratory, high-energy colliders and high-intensity experiments - as
the one explored in this contribution.   

From a theoretical point of view, (heavy) sterile fermions 
play an important role in several SM extensions which include 
well-motivated mechanisms of neutrino mass generation;
among these, one finds low-scale
realisations of the seesaw mechanism (including, for example,
low-scale type I and its variants, distinct realisations of the
Inverse Seesaw, as well as the Linear Seesaw), and their embedding
into larger frameworks, as for instance supersymmetrisations of the
SM, or Left-Right symmetric models.  

Before considering the contributions of these complete frameworks
(which typically call upon the heavy neutral leptons as a key
ingredient of the mechanism of neutrino mass generation) to the
distinct high-intensity muon observables previously described, 
it proves convenient - and insightful - to first carry a
phenomenological bottom-up approach. Without any formal assumption on 
the underlying mechanism of mass generation, 
the addition of a  
massive sterile state to the SM content 
allows to encode into a simple ``toy model'' the effects of 
a larger number of HNL states, possibly present in complete models.

\subsection{Bottom-up approach: $3+n_S$ toy models}
The ``toy models'' strongly rely on the assumption of having 
uncorrelated neutrino masses and leptonic mixings (or in other words,
that one does not consider a specific mechanism of $\nu$ mass
generation, for instance a seesaw). The model is described by a small
set of physical parameters, which include the masses of the 
$3$ mostly active light neutrinos, the
masses of the (mostly sterile) heavy neutral leptons, 
and finally the mixing angles and
the CP-violating phases encoded in the mixing matrix which relates the
physical neutrino to the weak interaction basis; for $n_S$ additional
neutral leptons, the matrix $U$ can be parametrised by $(3+n_S)(2+n_S)/2$ 
rotation angles, 
$(2+n_S)(1+n_S)/2$ Dirac phases and $2+n_S$ Majorana 
phases\footnote{In the case of a complete model of neutrino mass
  generation, the neutrino masses and
the   $(3+n_s)\times (3+n_s)$  lepton mixing matrix $U_{3+n_s}$ would
be formally derived   
from the diagonalisation of the full  $(3+n_S)\times (3+n_S)$ neutrino
mass matrix and thus be related; it is important to emphasise that in
such a case the model must necessarily account 
for $\nu$ oscillation data.}. 
For instance, in the simplest case where $n_S=1$ (the ``$3+1$'' model), 
the matrix $U_4$ can be constructed as follows
\begin{equation}\label{eq:para1}
 U_4 \,=\, R_{34}(\theta_{34},\delta_{43}) \cdot R_{24}(\theta_{24})
 \cdot R_{14}(\theta_{14},\delta_{41})  
\cdot \tilde{U} \cdot 
\text{diag}\left(1,e^{i\varphi_2},e^{i\varphi_3},e^{i\varphi_4}\right), 
\end{equation}
in which the Majorana CP-violating phases are factorised
in the last term. In the above, 
$R_{i j}$ is a unitary rotation matrix describing the mixing 
between $i$ and $j$
generations, parametrised in terms of the 
mixing angle $\theta_{ij}$ and of the Dirac CP-violating
phase $\delta_{ij}$. For example, $R_{14}$ can be cast as 
\begin{eqnarray}  
R_{14} &=& \left( 
\begin{array}{cccc}
\cos \theta_{14} & 0 & 0 & \sin \theta_{14}\, e^{-i \delta_{14}} \\
0 & 1 & 0 & 0 \\
0 & 0 & 1 & 0 \\
- \sin \theta_{14} \,e^{i \delta_{14}} & 0 & 0 &\cos \theta_{14} \\
\end{array}%
\right) \, .
\end{eqnarray} 
In the above, 
$\tilde{U}$ is a $4\times 4$  
matrix whose upper $3\times 3$ block encodes the mixing among
the left-handed leptons, and 
includes the ``standard'' Dirac CP phase. 
In the case in which the HNL decouples, this
sub-matrix would correspond to the usual unitary PMNS lepton mixing
matrix, ${U}_\text{PMNS}$. 
In the case of $n_S=2$, the definition of $U$
given in Eq.~(\ref{eq:para1}) can be extended  as
\begin{equation}
U_5\,=\,R_{45} \cdot R_{35} \cdot R_{25} \cdot R_{15} \cdot 
R_{34} \cdot R_{24} \cdot R_{14} \cdot R_{23} \cdot R_{13} \cdot R_{12}
\,\cdot\text{diag}\left(1,e^{i\varphi_2},e^{i\varphi_3},e^{i\varphi_4},
e^{i\varphi_5}\right)\,.
\label{eq:mixing}
\end{equation}

\subsubsection{Flavour conserving  observables}
\begin{figure}[t]
\begin{center}
\raisebox{3mm}{\includegraphics[width=0.45\textwidth]{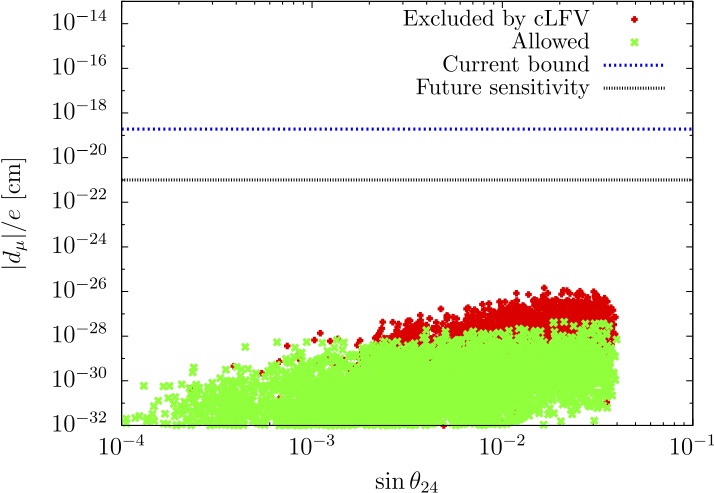}} 
\raisebox{-3mm}{\includegraphics[width=0.53\textwidth]{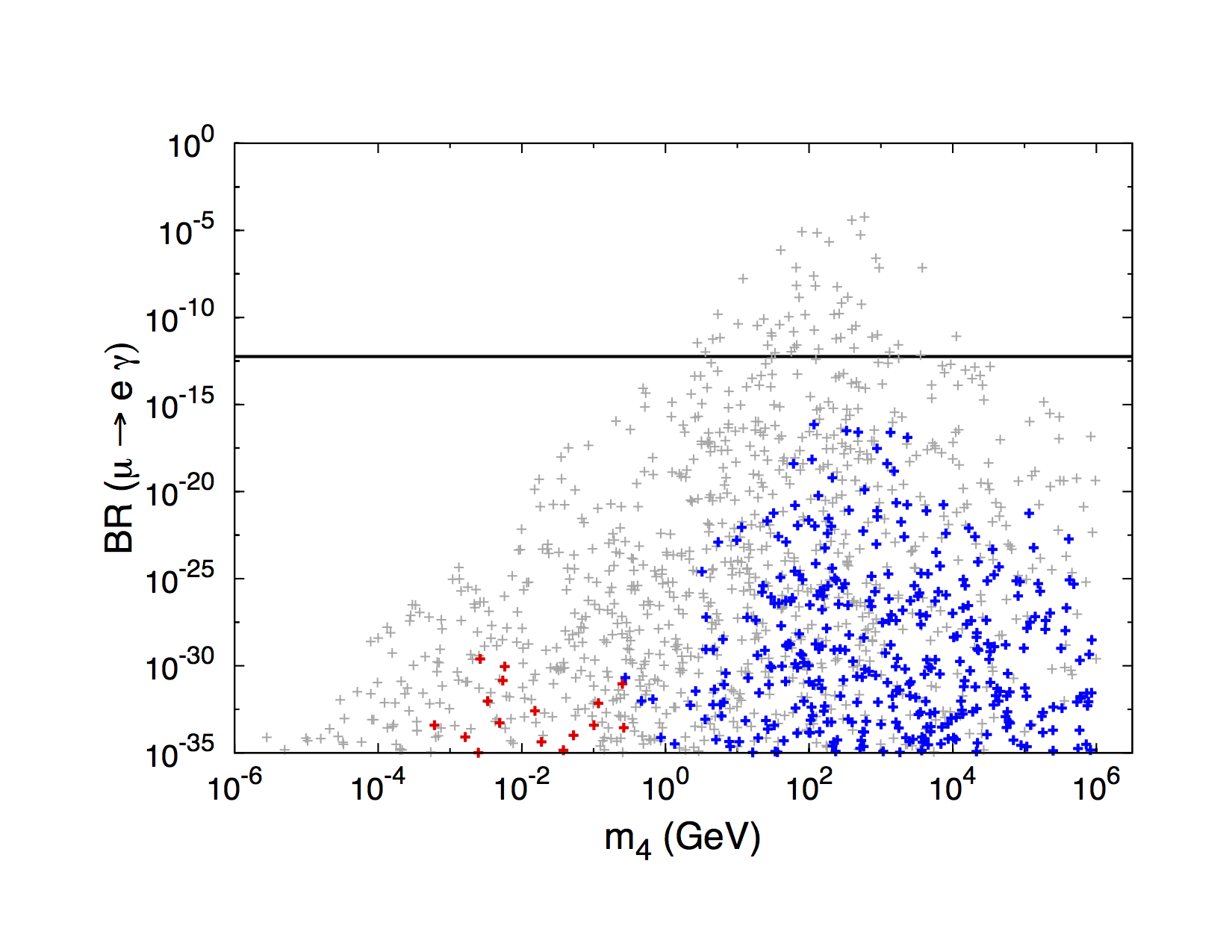}}  
\vspace*{-3mm}
\caption{Contributions to the muon EDM in a ``$3+2$''  
model as a function of $\theta_{24}$ (left panel); 
blue and black lines respectively denote 
the current upper bounds and future experimental sensitivity. 
From~\cite{Abada:2015trh}, reproduced with permission from the Authors.
On the right, BR($\mu \to e \gamma$) as a function of $m_4$;
grey points correspond to the violation of at least one experimental
bound and the horizontal line the current MEG bound.
} 
\label{fig:num_edm}
\end{center}
\end{figure}

As mentioned before, if the HNL are Majorana particles, 
they can have an impact regarding LNV  $0\nu2\beta$ decays, since the
new contributions to the effective mass can translate into enlarged
ranges for $m_{ee}$. The experimental implications are striking, 
given that the interpretation of a future signal can no longer be
associated to an inverted 
ordering of the light neutrino spectrum~\cite{Abada:2014nwa,Giunti:2015wnd}. 
Interestingly, the HNL can also be at the origin of contributions to a
distinct class of (lepton flavour conserving) observables, as is the
case of lepton EDMs, discussed in Section~\ref{sec:edm.g2}. 
The contributions of the HNL to the two-loop diagrams are dominated by
the terms associated with the new Majorana CP phases (the Dirac
contribution being in general sub-dominant), and become important
provided that there are at least two non-degenerate states, with
masses in the $[100\text{ GeV}, 100\text{ TeV}]$ range~\cite{Abada:2015trh};
the predictions obtained in a minimal ``$3+2$'' model are displayed in
Fig.~\ref{fig:num_edm} (left). As can be inferred, in such a 
minimal setup, one can have at best $|d_\mu|/e \sim
10^{-26}$~cm, which is far below the future sensitivity of J-PARC $g-2$/EDM
Collaboration~\cite{Saito:2012zz},
$|d_\mu|/e\sim10^{-21}~\mathrm{cm}$. 
By increasing the number of HNL ($n_S>2$) one could 
have an enhancement of a few orders of magnitude for the maximal values
of $|d_\mu|/e$; however, and since the charged lepton EDMs
approximately scale as 
$\frac{|d_e|}{m_e}\sim\frac{|d_\mu|}{m_\mu}\sim
\frac{|d_\tau|}{m_\tau}$~\cite{Abada:2015trh}, any future 
observation of $\sim10^{-21}~\mathrm{cm}$ for the muon EDM must
necessarily be interpreted in the light of another new physics scenario. 

As mentioned before, sterile states can also have an impact on other
flavour-conserving observables, as is the case of the muon anomalous
magnetic moment. Once all the experimental constraints discussed in
Section~\ref{sec:constraints} are imposed, 
in generic ``$3+n_S$'' scenarios, 
the predicted value of the muon anomalous magnetic moment is 
found to be
$|\Delta{a}_\mu|\lesssim 10^{-12}$ for $|U_{\mu
i}|^2\sim10^{-3}$ (with $i \geq 4$)~\cite{Abada:2014nwa}, 
and thus additional
contributions  to the anomalous magnetic moment are still required. 

\subsubsection{cLFV observables}
\begin{figure}
\begin{center}
\begin{tabular}{cc}
\includegraphics[width=0.49\textwidth]{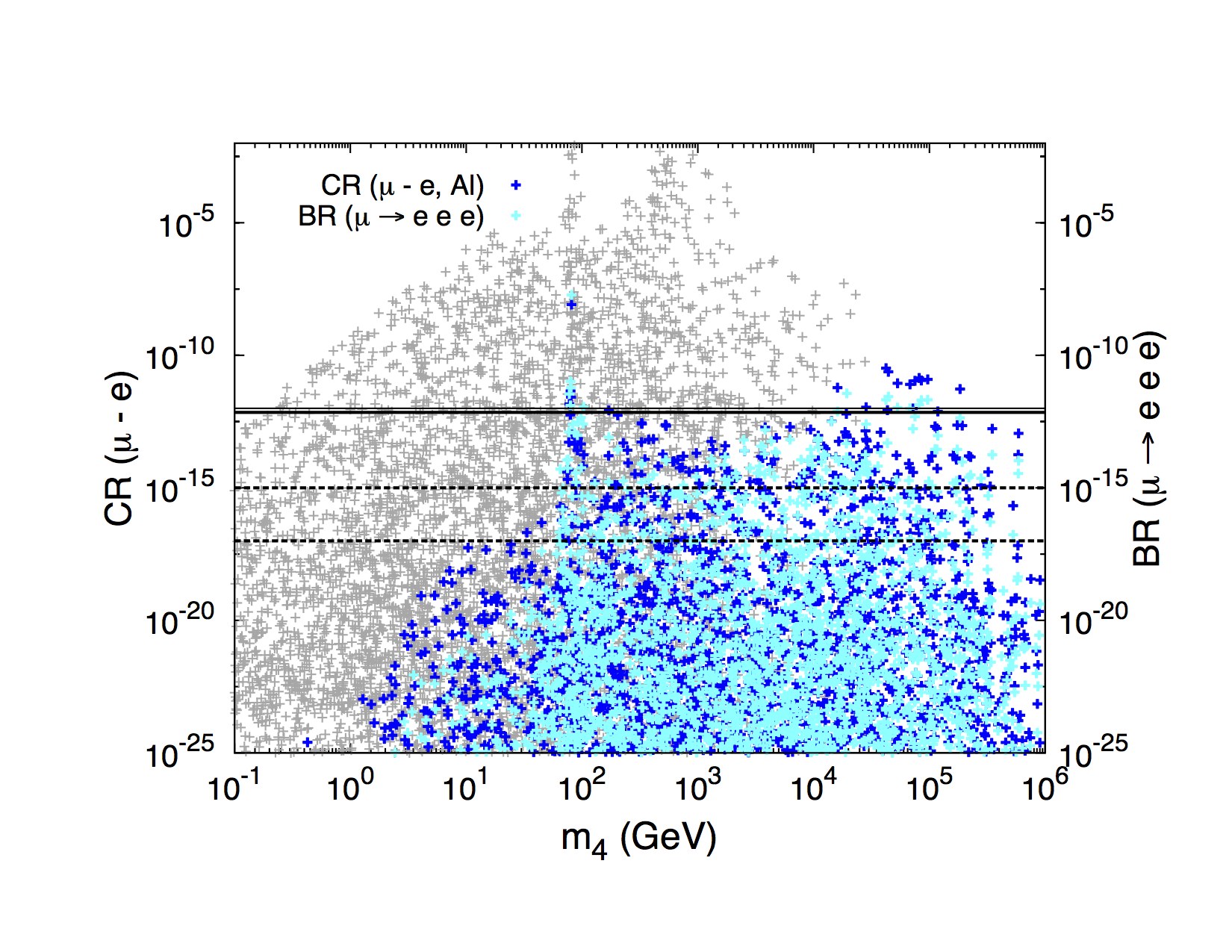} 
\hspace*{-0mm}&\hspace*{-0mm}
\includegraphics[width=0.49\textwidth]{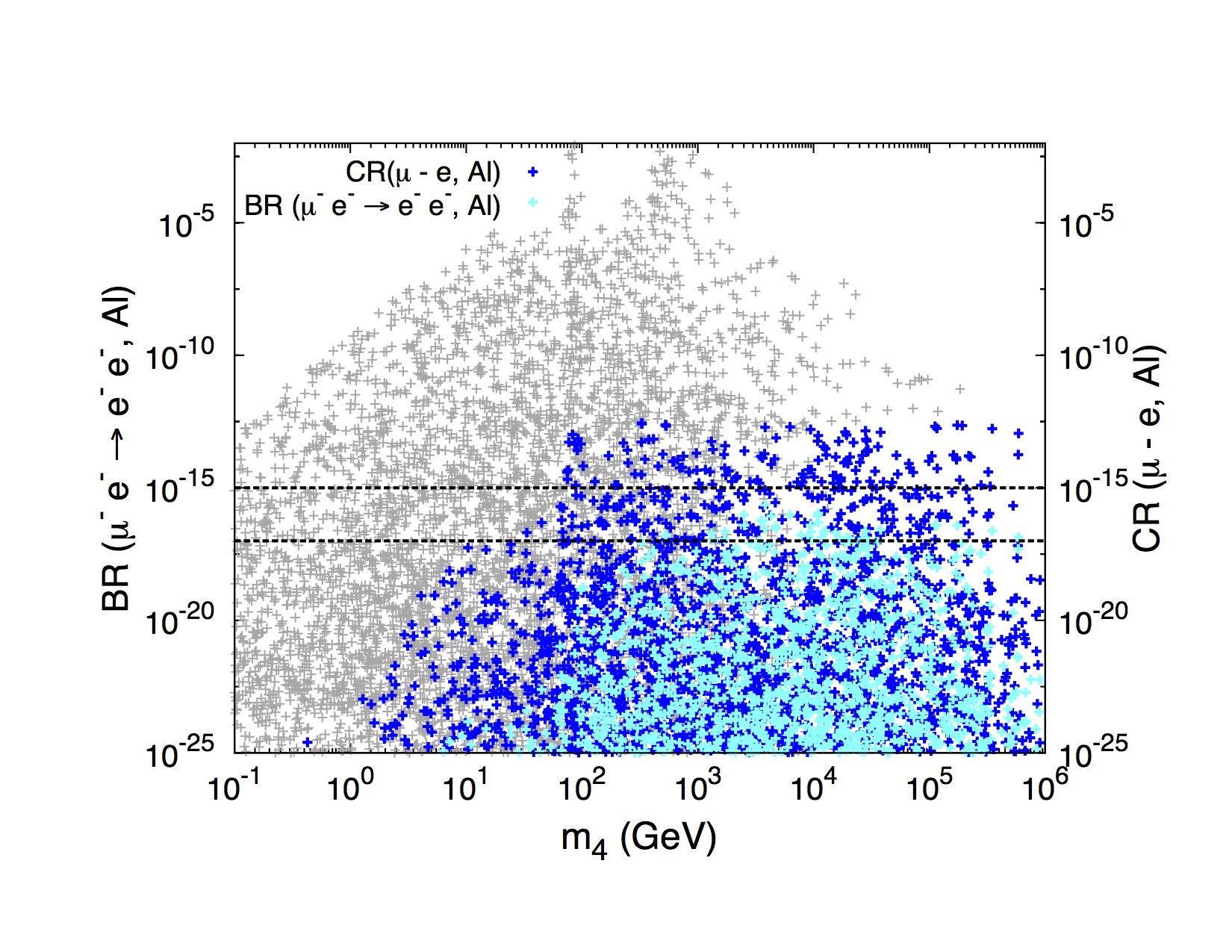} 
\end{tabular}
\end{center}
\vspace*{-6mm}
\caption{On the left, 
predictions for CR($\mu -e$, Al) and BR($\mu \to e e
  e$) as a function of $m_4$; the former is displayed in dark blue 
  (left axis), while the latter is depicted in cyan (right axis). 
A thick (thin) solid horizontal line denotes
the current experimental bound on the CR($\mu -e$,
Au)~\cite{Bertl:2006up} ($\mu \to e e e$ decays~\cite{Bellgardt:1987du}), 
while dashed lines correspond to future sensitivities to 
CR($\mu -e$, Al)~\cite{Kuno:2013mha,Bartoszek:2014mya}. 
On the right, BR($\mu^- e^- \to e^-e^-$) (cyan, left axis) and 
CR($\mu-e $, Al) (dark blue, right axis) as a function of $m_4$;
dashed horizontal lines denote the (expected) 
future sensitivity of COMET to both observables.
 Both figures were obtained in the ``3+1'' model, and in both panels
grey points correspond to the violation of at least one experimental
bound (from~\cite{Abada:2015oba}, reproduced with permission from the
Authors).  
} 
\label{fig:EFF_CRmue.mu3e_m4:mue.ee}
\end{figure}

Muon cLFV channels are in general very sensitive probes to the
presence of sterile fermions, in particular HNL with masses above the
electroweak scale. Beginning with radiative muon decays, the
contributions to BR($\mu \to e\gamma$) can be very large, well above 
the current bounds, as can be confirmed from the right panel of
Fig.~\ref{fig:num_edm}; however, these regimes are already excluded by 
other experimental constraints - in particular they are 
in conflict with bounds
arising from other cLFV muon channels, as is the case of 3-body decays
and $\mu-e$ conversion in Nuclei. 
The contributions of the HNL (obtained in a simple ``$3+1$'' extension) to
these two observables are 
displayed in the left panel of
Fig.~\ref{fig:EFF_CRmue.mu3e_m4:mue.ee}, as a function of the mass of
the heavy, mostly sterile, state, $m_4$. Especially for $m_4 \gtrsim
M_Z$, one can verify that the contributions are sizeable, within the
sensitivity of future $\mu-e$ conversion dedicated facilities (Mu2e
and COMET) and of Mu3e.

In the $\mu -e$ sector, 
neutrinoless conversion in nuclei (Aluminium) appears to be the
  cLFV observable offering the most
promising experimental prospects; nevertheless, 
the Coulomb-enhanced decay of a muonic
atom into a pair of electrons might prove to be also very competitive,
especially for heavy target nuclei (such as Lead or Uranium), since it
has been shown that the associated decay widths can be enhanced in
this case~\cite{Uesaka:2017yin}. 
Still in the framework of a minimal ``3+1'' model, the comparison of
the expected contributions to these observables can be found in the
right panel of Fig.~\ref{fig:EFF_CRmue.mu3e_m4:mue.ee}. For HNL states
heavier than the EW scale, both observables are within reach of COMET
(should the $\mu e \to e e $ decay be included in its Phase II
programme). 

It is interesting to notice that in the regime in which the 
mass of the HNL is heavier than the electroweak scale, the dominant
contributions to processes such as $\mu e \to eee$, $\mu-e$
conversion and $\mu e \to e e $ decays  
arise from $Z$-penguin exchange; this is at the source of a 
strong
correlation between the corresponding cLFV decays and the 
lepton flavour violating decays of the $Z$ boson, 
$Z \to \mu \ell$. Although marginal to the present discussion, we
notice that as pointed out in~\cite{Abada:2014cca}, the cLFV $Z$ decays
allow to probe $\mu-\tau$ flavour violation 
beyond the reach of Belle II.

\begin{figure}[h!]
\begin{tabular}{cc}
\includegraphics[width=0.49\textwidth]{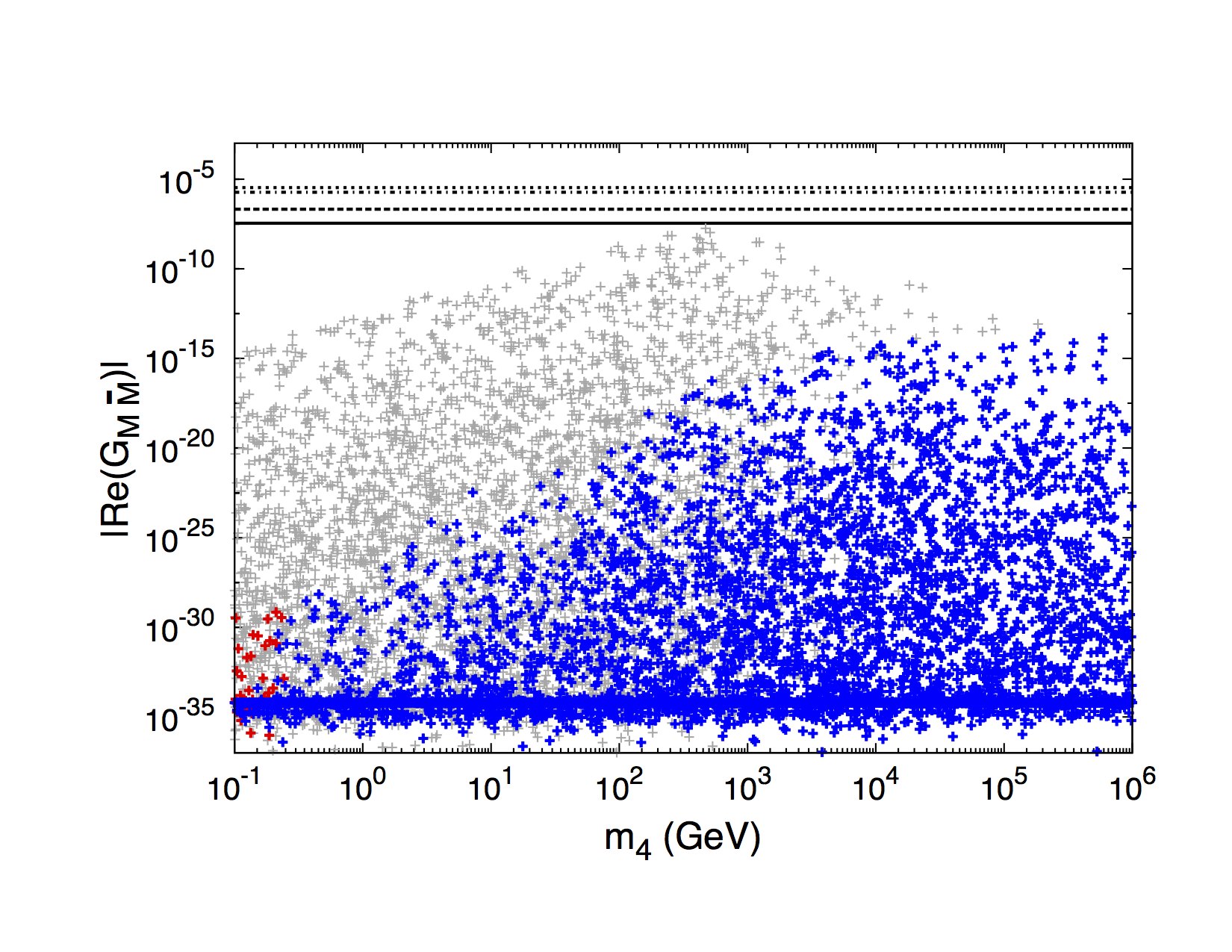}
\hspace*{0mm}&\hspace*{0mm}
\includegraphics[width=0.49\textwidth]{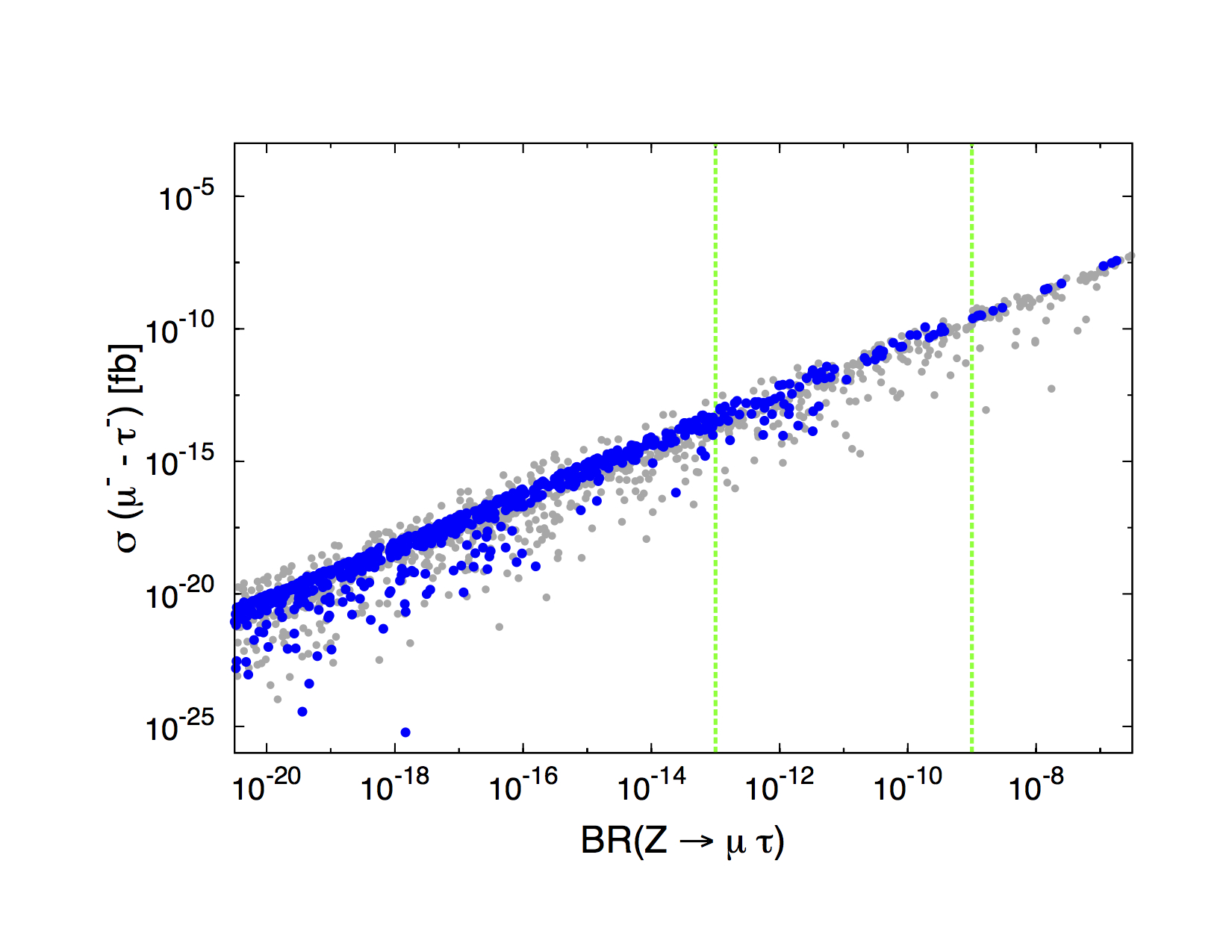}
\end{tabular}
\vspace*{-6mm}
\caption{On the left, 
effective coupling $G_\text{M$\rm \overline{M}$}$ ($\left
|\text{Re}\left(G_{\rm M\overline{M}} \right) \right|$) for Mu - $\rm
\overline{Mu}$ conversion 
as a function of $m_4$ (within the framework of a simple ``3+1 model''). 
Dark blue points are in agreement will all available bounds
(the horizontal lines denote 
the evolution of the experimental bounds and constraints); 
from~\cite{Abada:2015oba}, reproduced with permission from the
Authors.
On the right, correlation of cLFV in-flight $\sigma (\mu \to \tau)$  
vs. BR($Z \to \tau \mu$) in the ``3+1 model''; 
blue (grey) points denote allowed (excluded)
regimes, vertical green lines denote the future sensitivities; 
from~\cite{Abada:2016vzu}, reproduced with permission from the
Authors. 
In both panels, grey points correspond to the violation of at least
one experimental bound.}\label{fig:cLFV:Mu.inflight}
\end{figure}

Heavy sterile fermions can also lead to cLFV in association with the
Muonium system; the predictions for the contributions of an additional
sterile state (in a minimal ``3+1'' model) to Mu-$\rm \overline{Mu}$
oscillation are displayed in the left panel of
Fig.~\ref{fig:cLFV:Mu.inflight}; in view of the present experimental
roadmap, it remains unclear whether or not the HNL contribution could
be within future experimental reach. 
 
Finally, we comment on the prospects for cLFV in-flight conversion of
future intense muon beams, in particular focusing on the 
mode $\sigma (\mu \to \tau)$. 
Larger values of the cross-section, which could potentially be within
reach of a future Muon Collider (for nominal values of $10^{20}
\mu/\text{year}$), are in fact already excluded, as the associated
regimes (mass and mixings of the additional sterile) lead to values of
BR($\tau \to 3 \mu$) already in conflict with experimental 
bounds~\cite{Abada:2016vzu}. This is a consequence of having again
dominant contributions from $Z$-mediated penguins in both cases; this
is visible in the right panel of Fig.~\ref{fig:cLFV:Mu.inflight}, in
which we illustrate the prospects of $\sigma (\mu \to \tau)$ versus
the expected contributions to BR$(Z \to \mu \tau)$. The correlation of
the observables is clear, and further serves to illustrate the probing
power of flavour violating $Z$ decays (albeit at the high energy
frontier). 

\subsection{Complete NP frameworks and HNL: contributions to muon observables}

To conclude our brief overview, we thus consider a few illustrative
examples of complete SM extensions calling upon heavy neutral
fermions, focusing our attention 
on ``low-scale'' ($\lsim$ TeV) NP models. 
Other than low-scale
realisations of a type I seesaw, we will refer to 
many of its variations including well-motivated 
realisations such as the Inverse 
Seesaw (ISS)~\cite{Schechter:1980gr,Gronau:1984ct,Mohapatra:1986bd},  
the Linear Seesaw (LSS)~\cite{Barr:2003nn,Malinsky:2005bi}  and the 
$\nu$-MSM~\cite{Asaka:2005an,Asaka:2005pn,Shaposhnikov:2008pf}.
In addition, we also briefly comment on larger frameworks also
including HNL, and which have an important impact for the muon
observables here addressed. When relevant, we shall also discuss 
how the synergy of the distinct observables
might be instrumental in unveiling the NP model at work.

\subsubsection{Low-energy variants of Type I Seesaw}
The type I Seesaw relies in extending the SM content by at least two
additional ``heavy'' right-handed neutrinos.

\begin{figure}[h!]
\begin{tabular}{cc}
\raisebox{5mm}{
\includegraphics[width=60mm]{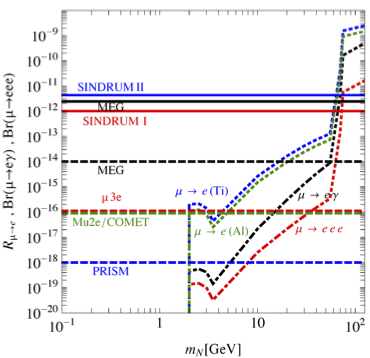}} 
&
\raisebox{-0mm}{
\includegraphics[width=92mm]{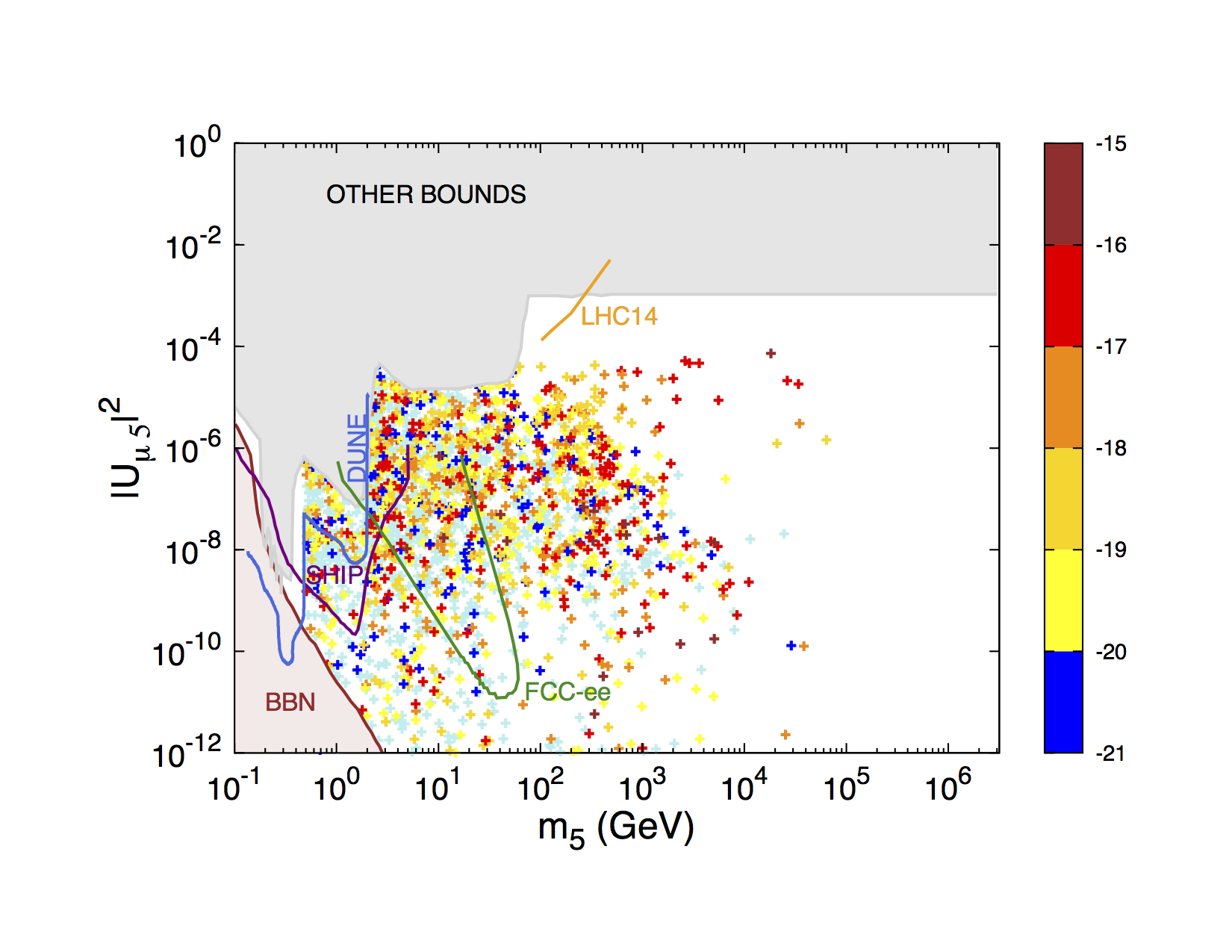}}
\end{tabular}
\vspace*{-6mm}
\caption{On the left, 
maximal allowed cLFV rates compatible with current searches
  in a low-scale seesaw; 
horizontal full (dashed) lines denote present (future) experimental 
sensitivity. From~\cite{Alonso:2012ji}, reproduced with permission from the
Authors.
On the right, logarithm of BR($\mu^- e^- \to e^- e^-$, Al),
  displayed on $(|U_{\mu 5}|^2, m_5)$ parameter
  space of a (3,3) ISS realisation; 
  the shaded surfaces correspond to the exclusion from BBN
  (rose) or from the violation of at least one experimental
  bound (grey), while solid lines delimit the expected sensitivity of
  several facilities (from~\cite{Abada:2015oba}, 
reproduced with permission from the Authors).
}\label{fig:seesaw}
\end{figure}

The light neutrino masses are given in terms of the Yukawa couplings
and of the RH neutrino mass matrix by the ``seesaw relation'', 
$m_\nu \sim -v^2 Y_\nu^\dagger M_R^{-1} Y_\nu$.
The {\it low-scale seesaw} (and its different variants) consists in a
realisation of a type I seesaw in which the 
(comparatively light) heavy mediators have non-negligible mixings with the
active neutrinos, and do not decouple.
Just as in the case of the
simple ``toy-models'' described in the previous section, the
modification of the leptonic currents can lead to 
contributions to numerous
observables~\cite{Dinh:2012bp,Alonso:2012ji}.
One such example - concerning contributions to cLFV muon radiative and
3-body decays, as well as $\mu-e$ conversion in nuclei -
can be found in the left panel of
Fig.~\ref{fig:seesaw}, in which the contributions to the distinct
observables (and the associated experimental bounds/future
sensitivities) are displayed as a function of the average seesaw
mediator mass. 

\vspace*{-4mm}

\bigskip
The {\it $\nu$MSM} consists in a specific low-energy realisation of a type I
seesaw, which aims at simultaneously
addressing the problems of neutrino mass generation, 
the BAU and providing a viable dark matter 
candidate~\cite{Asaka:2005an,Asaka:2005pn,Shaposhnikov:2008pf,
Canetti:2012kh}.  
The $\nu$MSM spectrum contains the three light (mostly
active) neutrinos, with masses given by a  type~I seesaw relation, as
well as three heavy states (with
masses $m_{\nu_{4-6}}$). In view of the model's goal to comply with 
the above requirements, 
the couplings and masses of the new states
are severely constrained. In particular,
and due to the smallness of the active-sterile mixings,
the expected contributions of the $\nu$MSM in what concerns cLFV
observables are found to lie beyond experimental
sensitivity. This has been discussed
in~\cite{Abada:2015oba,Abada:2016vzu}.   

\bigskip
Other than extending the SM by RH neutrinos, the {\it Inverse 
Seesaw}~\cite{Schechter:1980gr,Gronau:1984ct,Mohapatra:1986bd} calls
upon the introduction of additional sterile 
fermion\footnote{The minimal realisations of the Inverse Seesaw
  mechanism have ben discussed in~\cite{Abada:2014vea}.} 
states, $X$.
In the case of 3 generations of each,
the spectrum of the (3,3) ISS realisation
contains 6 heavy neutral fermions, which form 3 pseudo-Dirac pairs;
the smallness of the light (active) neutrino masses is explained  
by the suppresion due to the only source of LNV in the model
($\mu_X$), as given
by the following modified seesaw relation:
$m_\nu \approx \frac{(Y_\nu v)^2 }{(Y_\nu
  v)^2 + M_R^2} \mu_X$. This
allows for a theoretically natural model, in which one can have
sizeable Yukawa couplings for a comparatively light seesaw scale.  
On the right panel of Fig.~\ref{fig:seesaw} we illustrate the 
(3,3) ISS contributions to a muonic atom observable: 
the Coulomb enhanced decay into a pair of electrons, 
displaying the predictions for the corresponding BR
in terms of 
the mass of the lightest sterile state ($m_5$) and $|U_{\mu5}|^2$. As can be
seen, the contributions for these observables can be sizeable, well
within experimental reach. Particularly interesting is the fact that
these HNL states are within reach of future facilities such as DUNE, FCC-ee
and SHiP. Likewise, one expects important
contributions to other observables~\cite{Abada:2015oba}.

\bigskip
Another low-scale seesaw mechanism relying on an approximate
conservation of lepton number is the {\it Linear 
Seesaw}~\cite{Barr:2003nn,Malinsky:2005bi}. 
Similar to the case of the ISS, the Linear Seesaw also 
calls upon the addition of
two types of  fermionic singlets (RH neutrinos and other
sterile states) with opposite lepton number assignments.
However, in this case LNV is due to the
Yukawa couplings $Y'_\nu$ of the sterile states to the LH 
neutrinos. 
The resulting light neutrino masses are linearly 
dependent on these Yukawa couplings,
$m_\nu \approx 
(v Y_\nu) {(M_R^{-1})}^T {(v Y^\prime_\nu)}^T 
+  (v Y^\prime_\nu) {M_R}^{-1}\, (v Y_\nu)^T$.
The obtained spectrum in the mostly 
sterile sector is composed by pairs of pseudo-Dirac neutrinos 
(almost degenerate in mass) - a consequence of having the mass 
splittings determined by the small LNV couplings ($Y_\nu^\prime$), which are
also responsible for the suppression of the active neutrino masses. 
This is similar to what occurs in the ISS scenario, with which the LSS 
shares many phenomenological features (notice that distinctive
signatures can arise  
due to having two sources of flavour mixing, $Y_\nu$ and $Y_\nu^\prime$).

\subsubsection{Extended NP frameworks: LR models and SUSY}

\begin{figure}[h!]
\begin{tabular}{cc}
\includegraphics[width=55mm]{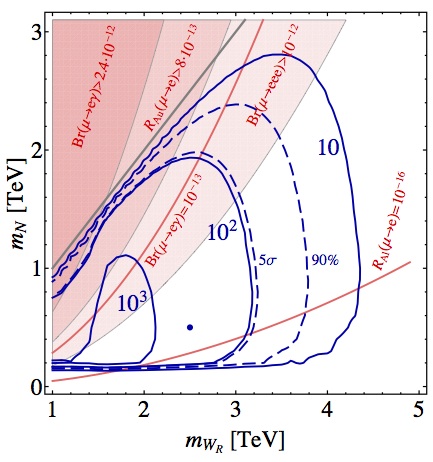}
\hspace*{2mm}&\hspace*{2mm}
\includegraphics[width=90mm]{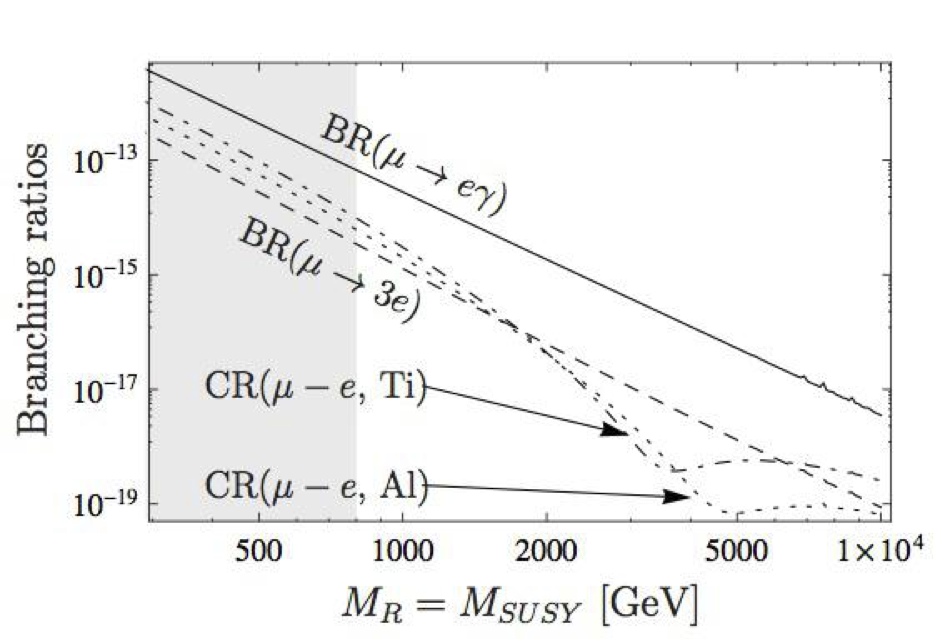}
\hspace*{5mm}
\end{tabular}
\caption{On the left, muon cLFV rates in LRSM: rose-shaded areas 
denote the corresponding experimental regimes (exclusion and future
sensitivity); solid lines
denote the number of events with a signature 
$e^\pm \mu^{\mp} + 2$~jets (no missing energy) 
at the LHC run 2 ($\sqrt s = 14$ and $\mathcal{L}=30\text{fb}^{-1}$), 
while the dashed ones define
regions with significances at $5\sigma$ (discovery) and 90\%
C.L. (exclusion). From~\cite{Das:2012ii}, and 
appearing also in~\cite{Deppisch:2012vj}  
(reproduced with permission from the Authors).
On the right, predictions for cLFV muon channels obtained in a SISS
realisation: BR($\mu \to e \gamma$), BR($\mu \to 3e$) and 
CR($\mu -e$, Al,Ti) as a function of $M_R=M_\text{SUSY}$.
The gray area roughly denotes regimes
excluded by direct LHC searches. From~\cite{Abada:2014kba}, 
reproduced with permission from the Authors.
}\label{fig:VL.LR}
\end{figure}

Restoring parity conservation in SM gauge interactions naturally leads
to models of NP which include HNL (right-handed neutrinos). 
In {\it Left-Right symmetric models}~\cite{LR:orig}, 
the SM gauge group is enlarged to 
{\small SU(3)$_c \times$SU(2)$_L \times$SU(2)$_R \times$U(1)$_{B-L}$},
and the particle content now includes, in addition to the RH
neutrinos, new $W_R$ and $Z_R$ bosons, as well as bi-doublet and
triplet (Higgs) bosons. 
Not only RH neutrinos are automatically incorporated
as part of an SU(2)$_R$ doublet 
upon realisation of the extended gauge group (and thus interacting
with the heavy right-handed bosons), but a hybrid type I-II seesaw
mechanism is at work in this class of models. 
Dirac neutrino mass terms arise from the interactions of the RH
neutrinos with the lepton doublets and the Higgs bi-doublets, while
Majorana mass terms are present for both left- and right-handed
neutral fermions, 
\begin{equation}\label{eq:LRSM:seesawI.II:massnu} 
M_\nu^\text{LR}\, =\, 
\left(
\begin{array}{cc}
M_L & m_D\\
m_D^T & M_R
\end{array}
\right)\, , \quad \quad 
\text{with} \quad \quad
\begin{array}{l}
m_D \, =\, Y_\nu\, \kappa \, +\,Y_\nu^\prime\, \kappa^\prime\,,\\
M_L\, =\, f_L\, v_L \,, \quad M_R\, = \,f_R\, v_R \,, 
\end{array}
\end{equation}
in which $Y^{(\prime)}$ and $f_{L,R}$ denote $3 \times 3$ complex Yukawa matrices
in flavour space; $\kappa$ and $\kappa^\prime$ are the vevs of the
Higgs bi-doublets, while $v_{L(R)}$ is the vev of 
the triplet $\Delta_{L(R)}$ (notice that 
$v_{R}$ is the vev responsible for 
breaking SU(2)$_R \times$U(1)$_{B-L}$ down
to U(1)$_Y$).
In the ``seesaw limit'' (i.e., for $|m_D| \ll |M_R|$),
block-diagonalisation of $M_\nu^\text{LR}$ in 
Eq.~(\ref{eq:LRSM:seesawI.II:massnu}) leads to a light
neutrino mass matrix of the form $
m_\nu = M_L - m_D  M_R^{-1} m_D^T$,
where both seesaw contributions are visible.
The new states (in particular the HNL and the
right-handed gauge bosons) lead to extensive contributions to many
muonic channels and, interestingly, to strong correlations between
high-intensity and high-energy cLFV and LNV observables (see, 
e.g.,~\cite{Cirigliano:2004mv,Das:2012ii,Deppisch:2012vj}.)
One such example (from~\cite{Das:2012ii,Deppisch:2012vj}) 
can be found on the left panel of
Fig.~\ref{fig:VL.LR}, in which the rose-shaded surfaces correspond to 
different regimes of contributions to $\mu \to e \gamma$, $\mu \to 3e$
and $\mu-e$ conversion in nuclei. Future sensitivities to 
$\mu-e$ conversion already allow to cover most of the parameter space
(here represented in $m_N,m_{W_R}$ plane), and further important
information can be inferred from cLFV decays at colliders: the blue
lines denote the number of events with a signature 
$e^\pm \mu^{\mp} + 2$~jets (no missing energy) at the LHC 
run 2 (assuming nominal values of $\sqrt s = 14$ and integrated
luminosity $\mathcal{L}=30\text{fb}^{-1}$), with 
dashed ones corresponding to $5\sigma$ significance (discovery) and 90\%
C.L. (exclusion).

\bigskip
The seesaw (in its distinct realisations) can be embedded in the
framework of supersymmetric (SUSY) extensions of the SM; in order to
render less severe the so-called ``SUSY CP and flavour problem'', the
seesaw is embedded in otherwise flavour and CP conserving SUSY models,
as is the case of the constrained Minimal SUSY SM (cMSSM). 
These BSM constructions offer many new contributions to cLFV
observables, as in general the different scales at work allow for
sizeable Yukawa couplings, and new - not excessively heavy - 
exotic mediators (sleptons and gauginos). 

For the ``standard'' {\it type I SUSY seesaw}, the right-handed neutrino
superfields (neutrinos and sneutrinos) are in general very heavy -
with masses $\mathcal{O}(10^{12-15})$~GeV. Interestingly,  
it has been emphasised that the synergy of
cLFV observables (among which the muonic ones here discussed) might
provide one of the best probes into the spectrum of the (extremely)
heavy neutrinos (see, for example~\cite{Antusch:2006vw,Figueiredo:2013tea}).

The supersymmetrisation of the ISS (in which case the HNL and their
SUSY partners are significantly lighter, closer to the TeV scale) - SISS - 
also leads to abundant signatures
in what concerns muon observables; a thorough study of several
observables (for different regimes, and taking into account distinct
contributions) was carried in~\cite{Abada:2014kba}. Here, we illustrate
the potential of the SISS via the contributions to several muon
channels as a function of the SUSY and seesaw scales ($M_R$), which
are displayed on the right panel of Fig.~\ref{fig:VL.LR}.

\section{Final remarks and discussion}
In the coming years, the High Intensity Frontier will offer many 
opportunities to explore particle and astroparticle physics. In addition
to testing some of the SM predictions, 
high-intensity experiments will open unique windows to probe New
Physics models. Many current tensions between the SM and observation 
are currently associated with the lepton sector: in addition to
neutrino oscillation data, several other observables (as is the case of
the muon anomalous magnetic moment) call for NP ingredients. 
As common ingredient to many BSM constructions, neutral leptons
(such as right-handed neutrinos, or other sterile fermions) play a key
role in many mechanisms of neutrino mass generation. 
The
motivations for these states are extensive: neutral fermions with
masses in the GeV-TeV range (``heavy neutral leptons'') are
particularly appealing, as in addition to a possible role in light
neutrino mass generation they might induce significant contributions
to many high intensity observables, such as cLFV, LNV or contributions
to lepton dipole moments, which can be searched for with the advent of
intense muon beams.  

In this small overview, we have discussed the contributions of HNL to
several observables which can be studied in 
high-intensity muon experiments. We have illustrated the potential of
the heavy sterile states via two complementary approaches:
considering ad-hoc ``$3+n_S$'' toy models 
and well-motivated
appealing NP models (seesaw mechanisms, LR models, supersymmetric
extensions of the SM, ...). As seen from our discussion, the new states can
easily give rise to significant contributions to many observables
well within experimental sensitivity (in fact, some of the current bounds 
already heavily constraining the associated new degrees of
freedom). Given their elusive nature, high-intensity muon
observables might be a unique probe of SM extensions via additional
heavy neutral leptons.

\section*{Acknowledgments}
We acknowledge partial support from the European Union Horizon 2020
research and innovation programme under the Marie Sk{\l}odowska-Curie: RISE
InvisiblesPlus (grant agreement No 690575)  and 
the ITN Elusives (grant agreement No 674896). 
We thank R. Bernstein and Y. Kuno for fruitful discussions.

\appendix
\section{Relevant form factors and loop functions for muon observables}
\label{app:A}
We collect in the Appendix the most relevant expressions for the
computation of the observables discussed in
Section~\ref{sec:muon.observables}; 
as mentioned in the text, the discussion is generic, and the formulae
here summarised hold for scenarios with
additional $n_S$ singlet neutrinos. 

\subsection{cLFV form factors}\label{appendix.form}
Written in a very compact way (and 
for simplicity for the case of $\ell_i= \mu$, $\ell_j= e$), 
we list below the relevant form
factors for the cLFV transitions and decays considered in 
Section~\ref{sec:muon.observables}, and refer
to~\cite{Alonso:2012ji,Ilakovac:1994kj,Ma:1979px,Gronau:1984ct} (or to
the summary in the Appendix A of~\cite{Abada:2015oba}) 
for a detailed discussion.  

\begin{eqnarray}
\label{eq:FF}
G^{\mu e }_\gamma &=& \sum_{j=1}^{3 + n_S} {U}_{ej}\,{U}^*_{\mu
  j} \,G_\gamma(x_j)\,,  \nonumber\label{Ggammamue} \\ 
F^{\mu e }_\gamma &=& \sum_{j=1}^{3 + n_S} {U}_{ej}\,{U}^*_{\mu
  j} \,F_\gamma(x_j)\,, \nonumber \label{Fmue}\\ 
F^{\mu e }_Z &=& \sum_{j,k=1}^{3 + n_S} {U}_{ej}\,{U}^*_{\mu k}
\left(\delta_{jk} \,F_Z(x_j) + {\bf C}_{jk}\, G_Z(x_j,x_k) + {\bf
  C}^*_{jk}\, H_Z(x_j,x_k)   \right)\,,  \\ 
F^{\mu eee}_{\rm Box}&=&  \sum_{j,k=1}^{3 + n_S}{U}_{e j}\, 
{U}^*_{\mu k}\left({U}_{e j}\,{U}^*_{ek}\,G_{\rm
  Box}(x_j,x_k)-2\,{U}^*_{e j}\,{U}_{e k}\,F_{\rm
  XBox}(x_j,x_k)\right)\,,   \\  
F^{\mu e uu}_{\rm Box}&=&\sum_{j=1}^{3 + n_S}\sum_{d_\alpha=d,s,b}
 {U}_{ej}\,{U}^*_{\mu j} \,V_{u d_\alpha} \,V^*_{u d_\alpha} \,F_{\rm
   Box}(x_j,x_{d_\alpha})\,,    \\
 F^{\mu e dd}_{\rm Box}&=&  \sum_{j=1}^{3 + n_S}\sum_{u_\alpha=u,c,t}
 {U}_{ej}\,{U}^*_{\mu j} \,V_{d  u_\alpha } \,V^*_{d u_\alpha} \,F_{\rm
   XBox}(x_j,x_{u_\alpha})\,. 
\label{Fmueee}  
\end{eqnarray}
The loop-functions entering in the above expressions can
be found in the following section (Appendix~\ref{appendix.loop}).

\subsection{Loop functions}\label{appendix.loop}
In what follows, we collect the most relevant loop functions 
involved in the computation of the observables detailed in 
Section~\ref{sec:muon.observables}.

\begin{equation}\label{eq:FM}
F_M\left(x\right)\,=\,
\frac{10-43x+78x^2-49x^3+4x^4+18x^3\ln{x}}{3\left(1-x\right)^4}.
\end{equation}

\begin{equation}\label{eq:steriles:Gloop}
G_\gamma(x) \, =\,
-\frac{2x^3+5 x^2 -x}{4\,(1-x)^3}
- \frac{3x^3}{2\,(1-x^4)}\, \ln(x)\,.
\end{equation}

\begin{align}
F_Z(x)&= -\frac{5x}{2(1-x)}-\frac{5x^2}{2(1-x)^2}\ln x \, , \nonumber
\\  
G_Z(x,y)&= -\frac{1}{2(x-y)}\left[	\frac{x^2(1-y)}{1-x}\ln x -
  \frac{y^2(1-x)}{1-y}\ln y	\right]\, ,  \nonumber \\ 
H_Z(x,y)&=  \frac{\sqrt{xy}}{4(x-y)}\left[	\frac{x^2-4x}{1-x}\ln
  x - \frac{y^2-4y}{1-y}\ln y	\right] \, , \nonumber \\ 
F_\gamma(x)&= 	\frac{x(7x^2-x-12)}{12(1-x)^3} -
\frac{x^2(x^2-10x+12)}{6(1-x)^4} \ln x	\, , \nonumber \\ 
G_\gamma(x)&=    -\frac{x(2x^2+5x-1)}{4(1-x)^3} -
\frac{3x^3}{2(1-x)^4} \ln x \, ,\label{Ggamma}  \nonumber \\	
F_{\rm Box}(x, y) &= \frac{1}{x - y} \bigg\{
\left(4+\frac{x  
y}{4}\right)\left[\frac{1}{1-x}+\frac{x^2}{(1-x)^2}\ln
x\right] - 2x 
y\left[\frac{1}{1-x}+\frac{x}{(1-x)^2}\ln
x\right] -(x\to y)\bigg\} \,,\nonumber \\
F_{\rm XBox}(x, y) &= \frac{-1}{x - y} \bigg\{
\left(1+\frac{x
y}{4}\right)\left[\frac{1}{1-x}+\frac{x^2}{(1-x)^2}\ln
x\right] - 2x
y\left[\frac{1}{1-x}+\frac{x}{(1-x)^2}\ln
x\right] -(x\to y)\bigg\} \,. 			
 \end{align}
In the limit of light masses ($x\ll 1$) and/or degenerate propagators
($x=y$), one has 
\begin{align}
F_Z(x) &  \xrightarrow[x\ll 1]{}    -\frac{5x}{2} \,,\nonumber\\
G_Z(x,x ) &= {} -\left[x (-1 + x - 2 \ln x)/(2 (x -1)) \right]]\, , \,\,
G_Z(x,x)  \xrightarrow[x\ll 1]{}  -\frac{1}{2} x \ln x \,, \nonumber\\
H_Z(x,x ) & = {} - \left[ \sqrt{x^2} (4 - 5x + x^2 + (4 - 2x + x^2)\ln
  x)/(4(x - 1)^2) \right] \, , \nonumber\\ 
F_\gamma(x ) & \xrightarrow[x\ll 1]{}  -x \,, \nonumber\\
G_\gamma(x )  & \xrightarrow[x\ll 1]{}  \frac{x}{4}\, \nonumber\\
 F_{\rm Box}(x,x)	 &= \left[\left(-16 + 31x^2 - 16x^3 + x^4 +
   2x(-16 + 4x + 3x^2) \ln x \right)/\left(4(-1 + x)^3\right) \right]
 \, , \nonumber\\ 
F_{\rm XBox}(x,x )& = \left[ (-4 + 19 x^2 - 16 x^3 + x^4 + 2x (-4  4 x
  + 3x^2) \ln x)/(4(x - 1)^3) \right] \, .\label{limitval2} 
\end{align}

Finally, for the Muonium system, one has
\begin{equation}
G_{\rm {Muonium}}(x_i, x_j) =  x_i x_j \left(
\frac{J(x_i)-J(x_j)}{x_i-x_j} \right)\,, 
\end{equation}
where 
\begin{equation}
J(x) = \frac{ (x^2-8x+4)}{4(1-x)^2} \ln x-\frac{3}{4}\frac{1}{(1-x)}.
\end{equation}
In the degenerate case, in which $x_i=x_j=x$, $ G_{\rm Muonium}$ is
given by
\begin{equation}
 G_{\rm Muonium}(x)  =  \frac{x^3 -11x^2
   +4x}{4(1-x)^2}-\frac{3x^3}{2(1-x)^3} \ln x\,. 
\end{equation}

\end{document}